\documentclass[manuscript,review=false,anonymous=false]{acmart} 

\usepackage{enumitem}

\setcopyright{rightsretained}
\copyrightyear{2026}
\acmYear{2026}
\acmDOI{XXXXXXX.XXXXXXX}

\acmConference[FAccT '26]{The 2026 ACM Conference on Fairness, Accountability, and Transparency}{June 27--30, 2026}{Montr\'{e}al, Canada}
\acmISBN{XXX}

\title{Job Anxiety in Post-Secondary Computer Science Students Caused by Artificial Intelligence}

\author{Daniyaal Farooqi}
\authornote{Both authors contributed equally to this research.}
\affiliation{%
  \institution{University of Toronto}
  \city{Toronto}
  \state{ON}
  \country{Canada}}
\email{daniyaal@cs.toronto.edu}

\author{Gavin Pu}
\authornotemark[1]
\affiliation{%
  \institution{University of Toronto}
  \city{Toronto}
  \state{ON}
  \country{Canada}}
\email{gpu@cs.toronto.edu}

\author{Shreyasha Paudel}
\affiliation{%
  \institution{University of Toronto}
  \city{Toronto}
  \state{ON}
  \country{Canada}}
\email{shreyasha.paudel@mail.utoronto.ca}

\author{Sharifa Sultana}
\affiliation{%
  \institution{University of Illinois Urbana-Champaign}
  \city{Champaign}
  \state{IL}
  \country{USA}}
\email{sharifas@illinois.edu}

\author{Syed Ishtiaque Ahmed}
\affiliation{%
  \institution{University of Toronto}
  \city{Toronto}
  \state{ON}
  \country{Canada}}
\email{ishtiaque@cs.toronto.edu}

\date{December 2025}

\begin{document}

\begin{abstract}
    The emerging widespread usage of AI has led to industry adoption to improve efficiency and increase earnings. However, a major consequence of this is AI displacing employees from their jobs, leading to feelings of job insecurity and uncertainty. This is especially true for computer science students preparing to enter the workforce. To investigate this, we performed semi-structured interviews with (n = 25) students across computer science undergraduate and graduate programs at \anon[a Canadian university]{the University of Toronto} to determine the extent of job replacement anxiety. Through thematic analysis, it was determined that computer science students indeed face stress and anxiety from AI displacement of jobs, leading to different strategies of managing pressure. Subfields such as software engineering and web development are strongly believed to be vulnerable to displacement, while specialized subfields like quantum computing and AI research are deemed more secure. Many students feel compelled to upskill by using more AI technologies, taking AI courses, and specializing in AI through graduate school. Some students also reskill by pursuing other fields of study seen as less vulnerable to AI displacement. Finally, international students experience additional job replacement anxiety because of pressure to secure permanent residence. Implications of these findings include feelings of low security in computer science careers, oversaturation of computer science students pursuing AI, and potential dissuasion of future university students from pursuing computer science.
\end{abstract}




\maketitle

% https://facctconference.org/2026/authorguide.html

\section{Introduction}
% \todo{IA: You need to "sensitize" the Introduction section more with recent anecdotes, stats, and/or news that talks about AI and unemployment.}
\label{sec:introduction}

The advancement of artificial intelligence (AI) technology in the last few years has urged workers and employers to evaluate its effects on productivity. The ability of generative AI models to create new text, images, video, audio, and other modalities can speed up industrial task completion \cite{golda_privacy_2024,denning_artificial_2025}. In particular, AI has been shown to boost productivity in programmers; using an LLM designed to assist programming teams resulted in a 55\% increase in lines of code written \cite{gambacorta_generative_2024}, and the number of accepted GitHub Copilot line completions was found to correlate with perceived productivity \cite{ziegler_measuring_2024}. Through the ability to generate new content of economic value, AI can boost labor output, thereby allowing corporations to accumulate profits faster \cite{sestino_leveraging_2022,fox_artificial_2024}.

However, the rising use of AI tools and services in corporate settings has raised ethical concerns, one of which is employees' unease on whether their own jobs may be supplanted by AI \cite{denning_artificial_2025,denning_dilemmas_2020}. By 2030, 400 to 800 million jobs that existed in 2017 may be displaced by automation \cite{manyika_jobs_2017}. While the World Economic Forum projects that 170 million jobs will be created between 2025 and 2030, another 92 million existing jobs are expected to be displaced, and 39\% of workers' existing skill sets will be transformed or outdated by then \cite{forum_future_2025}. This suggests that a significant amount of upskilling and reskilling is necessary to remain in the workforce \cite{tomar_future_2024}. The threat of AI dispossessing professionals of their work has led to feelings of anxiety surrounding future career prospects. As in previous literature, we refer to this as ``AI job replacement anxiety" \cite{wang_what_2022}.

AI job replacement anxiety among professionals is not unfounded. Its unprecedented ability to generate original content and automate cognitive tasks poses a real threat to current workers, who are overlooked and disregarded in favor of automation \cite{norris_ai_2024,liu_what_2025,constantinides_ai_2025}. Large tech firms like Amazon, Google, Microsoft, and Salesforce have already laid off workers and replaced them with AI, with Salesforce alone eliminating 4,000 customer support positions \cite{osullivan_these_2025}. Jobs vulnerable to AI displacement span across a diverse range of fields and industries including software engineering, information technology, higher education, manufacturing, customer service, and medicine \cite{ahmed_artificial_2025,sharma_psychological_2025,memarian_employment_2024,george_chatgpt_2023,sallam_anxiety_2024}. Even post-secondary students, who are preparing to enter the workforce at a time when AI has become a fixture in society, have frequently mentioned job replacement and unemployment as among the most pressing ethical concerns of AI; students from Hong Kong have directly expressed fears that generative AI may replace current positions of interest, cause them to lose jobs in the future, and widen income disparities \cite{ghotbi_attitude_2022,chan_students_2023}. Yet despite the importance of understanding post-secondary student perspectives on how AI may impact their future careers, existing research has mainly focused on the threat of AI to the career stability of current professionals \cite{ji_current_2024, sun_research_2021, kam_what_2025, ahmed_artificial_2025}.

% Existing research has focused on the threat of AI to career stability of current professionals \cite{ji_current_2024, sun_research_2021, kam_what_2025, ahmed_artificial_2025}. However, there is less research concerning the perspective of post-secondary students, who are currently preparing to enter the workforce in a time when AI has become a fixture in society.

% \citet{wang_what_2022} explored AI job replacement anxiety in Taiwanese students of any major, but did not explore whether reactions were disproportionate across different demographic groups. While one study showed that female employment is threatened by AI \cite{sun_research_2021}, no studies showed that there is a greater impact, perceived or real, on female employment than male employment. Moreover, while there are several quantitative studies and literature reviews concerning students' viewpoints on AI's effects on jobs and employment \cite{wang_what_2022, rizun_employment_2024, sakib_chatgpt_2023, clos_investigating_2024}, there are no qualitative studies that explore this topic with in-depth interviews of affected students. Lastly, no empirical studies focus on computer science students alone.

% Expanded on the paragraph below a bit, will reread it later to see if I still think it makes sense - G
% In the context of our study, we define \textit{upskilling} as learning additional skills related to AI to keep up with industry trends and \textit{reskilling} as learning additional skills to transition to a different job or field.

As in previous literature, we define \textit{upskilling} as learning additional skills within one's current industry to keep up with emerging trends and \textit{reskilling} as learning additional skills outside of one's current industry to transition to a different career or field \cite{george_chatgpt_2023,cramarenco_impact_2023,morandini_impact_2023}. In the context of post-secondary computer science students facing pressure from AI, upskilling and reskilling may refer to learning new skills within the field of AI to keep pace with trends in industry, and learning skills unrelated to AI to pivot to a different discipline, respectively.

Our research seeks to explain the perspectives of post-secondary computer science students through interviews, exploring AI job replacement anxiety in greater depth than previous studies. To achieve our aims, we propose the following research questions:

% Removed this from the previous paragraph since the only demographic group we're comparing is domestic/international, but here for safekeeping for now
% and identifying whether levels of anxiety and views on future career prospects differ across demographic groups

\begin{quote}
\noindent \textbf{RQ1}: To what extent do computer science students experience AI job replacement anxiety?\\
\noindent \textbf{RQ2}: Which fields do computer science students feel discouraged from entering or perceive as vulnerable to displacement?\\
\noindent \textbf{RQ3}: Does AI job replacement anxiety encourage upskilling by becoming more adept at using AI tools and/or reskilling by changing fields or study?\\
\end{quote}

Our study reports that computer science students experience significant job replacement anxiety due to the displacement of entry-level positions. Furthermore, international students experience it to an even greater degree due to need to secure permanent residence. Software engineering and development, web development, and game development are seen as particularly vulnerable to displacement as LLMs are perceived by some students to be as adept at programming as junior developers. Meanwhile, specialized fields such as quantum computing are seen as less replaceable, although they often require graduate school. Finally, AI job replacement anxiety encourages upskilling by taking more AI courses and using AI more frequently. It also encourages reskilling by adopting a second major and gaining skills unrelated to computer science.
\section{Related Work}
\label{sec:related work}

\subsection{Job Displacement by AI}

The core concern of job displacement by AI can be attributed to economic incentives. Corporations are able to accumulate more profits if the cost of human labour were reduced via automation in the form of AI \cite{acemoglu_wrong_2020,constantinides_ai_2025,fox_artificial_2024,rehak_ai_2025}. Thus, concerns over the impact of AI on the labour market have been noted by researchers across various countries including the US, China, and India \cite{press_how_2024,zhang_ai-driven_2025,nigar_artificial_2025,sharma_psychological_2025}. While new employment opportunities may be created as AI skills gain demand \cite{dahlin_who_2024}, many more existing positions will close due to automation by AI, forcing workers to upskill or reskill to maintain career prospects \cite{kam_what_2025,pandya_artificial_2024,cramarenco_impact_2023}.

% Examples of job displacement within CS industries: software engineering, cybersecurity

Among subfields of computer science, software engineering and cybersecurity have been cited as two fields that have experienced AI job displacement. AI's ability to generate code, tests, documentation, and assets has made it compatible with accelerating the work of current software engineers \cite{kam_what_2025,malheiros_impact_2024}. Tasks within cybersecurity including analyzing and extracting information from system logs or detecting system access patterns are easily automated by AI \cite{nowrozy_gpts_2024}. Across both fields, however, the increasing usage of AI has led employees to question whether human developers may eventually become displaced. Roles in both fields involving repetitve tasks that do not require much insight or creativity, such as basic coding and manual testing, are particularly vulnerable to automation by AI \cite{ahmed_artificial_2025,nowrozy_gpts_2024}. Anxieties have emerged among junior developers over upskilling to use AI tools effectively rather than be replaced by them. As the development of AI may lead to higher demand for specialized skill sets, concerns also arise over whether the skill and income gaps between entry-level and senior employees will continue to widen \cite{ahmed_artificial_2025,kam_what_2025}.

\subsection{Effects of AI in Industry}

As technological advancements usually benefit skilled workers over unskilled workers, the wage gap between the skilled and unskilled workers may increase, thus widening the labour income gap \cite{ji_current_2024,frank_ai_2025}. Additionally, medium-skill positions are susceptible to displacement, with AI potentially converting them into unskilled workers due to AI automation of skilled labour \cite{ji_current_2024,kudoh_robots_2025}. Though AI may foster enough economic development in the long term to allow for upskilling, reskilling, and market adjustments, AI has still been projected to increase unemployment in the short term by replacing workers faster than both the emergence of new job positions and the development of new skills by current workers, who may not have enough time to adapt \cite{qin_artificial_2024,jung_scoping_2024}.

Professionals with resilient mindsets were found to be better at facing uncertainty over their job status due to AI \cite{chung_artificial_2025}. However, this uncertainty can also spread detrimental feelings and actions across employees, who perceive the risk of job loss. The possibility of AI displacing one's job may lead to feelings of unfulfillment, unimportance, or withdrawal; deviant behaviours such as taking longer breaks than allotted and working on personal instead of job-related tasks; and a higher chance of quitting and greater turnover rates \cite{chung_artificial_2025, dou_does_2023, ahmed_artificial_2025, sharma_psychological_2025,cordasco_self-esteem_2025}. While new job roles may be also be created as the growth of AI opens new opportunities, the threat posed by AI automation of certain jobs is still a major issue as career progressions could be impacted and individual economic security may be at risk \cite{tomar_future_2024}.

\subsection{Upskilling and Reskilling}

As AI is a rapidly growing technology, upskilling and reskilling become critical in determining whether workers will succeed \cite{cramarenco_impact_2023,li_reskilling_2024}. According to the World Economic Forum, 59\% of the world's workforce would need upskilling or reskilling by 2030, with AI and big data referenced as the two fastest-growing skills \cite{forum_future_2025}. AI's ability to automate widely applicable tasks like writing, teaching, translation, and data analysis reduces the need for workers to have these skills and may potentially lead to deskilling or outright replacement of workers who are not learning new skills \cite{yang_deskilling_2026}. Hence, lifelong learning in the form of upskilling and reskilling is becoming a necessity for the workforce. In particular, skills that emphasize innovation, creativity, critical thinking, and complex problem solving are among the most important for the current workforce to be prepared for the future and less vulnerable to automation \cite{li_reskilling_2024,shimray_ai_2025}.

% existing student reactions to AI
\subsection{Job Replacement Anxiety in Students}

\citet{wang_what_2022} investigated how students' ``AI learning anxiety" (pressure to upskill by taking AI courses) and AI job replacement anxiety affected intrinsic and extrinsic learning motivations. The authors made a distinction between debilitating anxiety, which interferes with performance, and facilitating anxiety, which improves performance. Ultimately, AI learning anxiety was categorized as debilitating anxiety, which negatively affects intrinsic and extrinsic motivation, though intrinsic motivation is affected more. Meanwhile, the authors categorized AI job replacement anxiety as facilitating anxiety, with a positive effect on extrinsic motivation.

Concerns surrounding AI job displacement have been affecting students \cite{sakib_chatgpt_2023,clos_investigating_2024}. While AI was perceived by some students to have minimal effects on unemployment and as a useful tool for career preparation \cite{ravselj_higher_2025,moran_can_2024}, other students expressed anxiety over future job replacement and falling behind due to AI \cite{chan_students_2023}. \citet{rizun_employment_2024} found that while students may see the current employment landscape as an opportunity rather than a risk, a substantial amount of students still reported feeling inadequately prepared to navigate it. Among specific fields, AI job replacement anxiety has been reported by students in computer science, social science, fine arts, and medicine \cite{chan_students_2023,wei_i_2025,sallam_anxiety_2024}. International students are often motivated to study abroad due to future career prospects, but may be more vulnerable to AI job displacement due to the additional challenges of language barriers and cultural unfamiliarity \cite{wang_exploring_2023}.

The effect of job replacement anxiety on students is significant since they are the future of their respective fields. However, the effect on computer science students specifically is hardly found in existing literature. Furthermore, while there are several quantitative studies and literature reviews concerning students' viewpoints on AI's effects on jobs and employment \cite{wang_what_2022,rizun_employment_2024, sakib_chatgpt_2023,clos_investigating_2024,chan_students_2023,ghotbi_attitude_2022}, there are no qualitative studies that explore this topic with in-depth interviews of affected students. We aim to address these gaps by exploring AI job replacement anxiety of students in Canada, where the unemployment rate of youths aged 15--24 (13.3\%) was over double the unemployment rate of adults aged 25--54 (6.0\%) by the end of 2025 \cite{statistics_canada_labour_2026}. Since almost one-third of post-secondary students in Canada are international \cite{statistics_canada_postsecondary_nodate}, we also aim to provide insight on whether international and domestic students experience different levels of AI job replacement anxiety.

% Finally, no studies were performed in Canada, and no studies explore whether international and domestic students experience different levels of AI job replacement anxiety.
\section{Methods and Data}
\label{sec:methods}

% \textcolor{red}{SS note: You guys can try following this way (see the table in the paper in the link, page 7, of presenting the participant details, https://dl.acm.org/doi/pdf/10.1145/3701201}

The study was conducted at \anon[a Canadian university]{the University of Toronto}, targeting undergraduate and graduate computer science students. The design and methods of the study were approved by \anon[the IRB of the authors' institution]{the University of Toronto's IRB}. To recruit student participants, we distributed flyers in \anon[the computer science building]{the Bahen Centre for Information Technology at the University of Toronto} and performed snowball sampling \cite{goodman_snowball_1961,biernacki_snowball_1981}.

Our main mode of data collection was 15-minute semi-structured interviews conducted by the main authors. Before the interviews, we sent out a pre-collection survey, which was accessible through a QR code on the flyer. The survey first described the details of the study and asked for informed consent from the participants. All personal information was optional aside from their name, email, degree type, year of study, and computer science program. Making this information mandatory was deemed necessary to ensure that all participants were enrolled in a computer science program at the \anon[Canadian university]{University of Toronto}.

In the optional questions, the survey asked for demographic information. This included their age, race or ethnicity, gender identity, first generation status, international student status, and self-identified disabilities. Second, we asked for information regarding the participant's student status. This included their focuses or concentrations, cumulative GPA, computer science subfields of interest or specialization, and plans after graduation. Third, we asked students to provide their level of agreement with 8 statements related to job anxiety caused by AI on a 7-point Likert scale to preliminarily gauge AI anxiety and strategies taken to mitigate it. These statements are available in Table \ref{table:likert}. Finally, the survey invited participants to a 15-minute online interview through the platform Arrangr, which if accepted sent an email to book a time slot for an interview over Zoom. See the \textbf{Appendices} for details on the Interview Protocol and how it was developed.

% \begin{table}[h!]
% \centering
% \caption{7-pt Likert scale survey questions}
% \begin{tabular}{|p{0.9\linewidth}|}
% \hline
% \begin{itemize}[leftmargin=1em]
%     \item \textit{I am stressed or anxious about my future career prospects because of AI}
%     \item \textit{I feel that my future career roles could be replaced or displaced because of AI}
%     \item \textit{I have upskilled or been motivated to upskill because of AI}
%     \item \textit{I have reskilled or been motivated to reskill because of AI}
%     \item \textit{I have selected courses I normally would not have considered taking because of AI}
%     \item \textit{I have decided against selecting courses I previously wanted to take because of AI}
%     \item \textit{I have considered leaving my current field of study or have left a field of study in the past because of AI}
%     \item \textit{I have felt discouraged from entering certain roles or industries I was once considering because of AI}
% \end{itemize}
% \\ \hline 
% \end{tabular}
% \end{table}

\subsubsection*{Note on Data Collection}
We collected information such as age, race or ethnicity, gender identity, first generation status, international student status, self-identified disabilities in case there were significant correlations with AI job replacement anxiety, but kept it optional in case students did not wish to provide this information. All students indicated whether they were an international student or domestic student, allowing us to make related insights. All information is kept confidential and was provided with informed consent.

Data was collected during interviews using the Zoom transcript and recordings, which we confirmed with every participant before the recording began. Finally, all interview participants were compensated with a \$10 CAD Amazon gift card.

% \subsection{Participants}

% The age range was 17 to 32. 18 participants identified as male, 6 identified as female, and one preferred not to say. We believe the gender disparity in our data was a result of a smaller female population in computer science education \cite{tereshchenko_why_2024}.

% 18 interview participants were enrolled in a Bachelor's program, while 7 were enrolled in a professional Master's program. From the sample of students pursuing Bachelor's degrees, we interviewed 5 first years, 4 second years, 2 third years, and 6 fourth years. From the sample of students pursuing Master's degrees, we interviewed 6 first years and 1 second year.

% 18 interview participants were enrolled in a Bachelor's program, while 7 were enrolled in \anon[a professional Master's program]{the Master's of Science in Applied Computing, a professional Master's program}. From the sample of students pursuing Bachelor's degrees, we interviewed 5 first years, 4 second years, 2 third years, and 6 fourth years. From the sample of students pursuing Master's degrees, we interviewed 6 first years and 1 second year.

% The cumulative GPA range was from 2.38 to 3.98 among undergraduates, and exclusively 4.00 among graduates. 

% 5 specialists, 8 majors but 'specialist' is a uoft giveaway
% From the undergraduates, there were 13 computer science majors, and 5 first-year admission stream students.

% Finally, 7 international students and 18 domestic students were interviewed.

\section{Results}
\label{sec:results}

Our sample consisted of computer science undergraduate and graduate students at \anon[a Canadian university]{the University of Toronto}. 28 participants filled out the survey, but only 25 participants were interviewed, as 3 survey participants did not attend. Of the participants who completed an interview, the age range was 17 to 32. As Table \ref{table:summary} shows, the majority of interview participants identified as male. We believe the gender disparity in our data was a result of a smaller female population in computer science education \cite{tereshchenko_why_2024}.

\begin{table}[ht]
\centering
\caption{Summary of interview participants}
\begin{tabular}{c c}
\toprule
\textbf{Gender} & Male (18), Female (6), Prefer not to say (1)\\
\midrule
\textbf{Degree and} & Bachelor's (18): Year 1 (5), Year 2 (4), Year 3 (2), Year 4 (7)\\
\textbf{Year} & Master's (7): Year 1 (6), Year 2 (1)\\
\midrule
\textbf{Domestic or International} & Domestic (18), International (7)\\
\bottomrule
\end{tabular}
\label{table:summary}
\end{table}

Through thematic analysis, the following themes were found in interviews. Students expressed varying degrees of AI job anxiety related to their use of AI or lack thereof. Some students saw AI use as necessary for learning the skills required to become competitive in a challenging job market while others were more reluctant to do so. Software engineering and web development were seen as fields vulnerable to displacement by AI. Upskilling is a common theme across many students who felt encouraged or pressured by the adoption of AI across many tech industries to learn more about AI. Reskilling is an additional strategy for students to feel more secure against AI displacement of jobs by pursuing additional fields that were seen as more specialized or less likely to fall to automation. Other concerns include additional pressures on international students to secure a job after graduation in order to remain in Canada.

\subsection{Extent of AI Job Replacement Anxiety}
\textit{Students experience varying degrees of job replacement anxiety, with most students feeling some extent of stress over AI displacement of jobs.}
% \textcolor{red}{SS Note: This table might actually be a horizontal bar-chart, right?}
\begin{table}[ht]
\centering
\caption{Average scores for 7-pt Likert scale survey questions}
\footnotesize
\begin{tabular}{p{0.80\linewidth} p{0.10\linewidth}}
\toprule
\textbf{Statement} & \textbf{Score (/7)} \\
\midrule
I am stressed or anxious about my future career prospects because of AI & 4.54\\
I feel that my future career roles could be replaced or displaced because of AI & 4.54\\
I have upskilled or been motivated to upskill because of AI & 4.82\\
I have reskilled or been motivated to reskill because of AI & 3.61\\
I have selected courses I normally would not have considered taking because of AI & 3.61\\
I have decided against selecting courses I previously wanted to take because of AI & 2.61\\
I have considered leaving my current field of study or have left a field of study in the past because of AI & 3.21\\
I have felt discouraged from entering certain roles or industries I was once considering because of AI & 4.29\\
\bottomrule
\end{tabular}
\label{table:likert}
\end{table}

\subsubsection{Survey Responses}

The survey participants agreed that they have stress or anxiety about their career prospects because of AI, and that they feel their roles could be displaced (4.54/7.00). They similarly agreed to having upskilled or been motivated to upskill (4.82/7.00) and to feeling discouraged from particular roles or industries (4.29/7.00). Students were neutral on reskilling (3.61/7.00) and selecting courses they normally would not have considered taking due to AI (3.61/7.00). Meanwhile, students mildly disagreed with considering leaving their field or study or leaving one in the past due to AI (3.21/7.00) and disagreed with deciding against selecting courses because of AI (2.61/7.00).

Overall, 23 out of 28 survey participants gave a 4/7 or higher for their stress about future career prospects because of AI, with 11 of them giving a 5/7 or higher. In terms of feeling that their future career roles could be displaced, 21 out of 28 participants gave a 4/7 or higher while 10 out of 28 participants gave a 5/7 or higher. This was further reiterated during interviews, where students gave a variety of reasons for their anxiety, which are discussed in the following subsections.

\subsubsection{Reasons for Anxiety}

During interviews, students justified their anxiety in different ways. Some students felt that they were unable to compete or keep up with AI and saw the job market as more difficult than before due to AI replacing jobs that new graduates would have applied to in the past, such as junior developer roles. Students mentioned that with the reduction of these roles, they may find it harder to secure employment after graduation. One undergraduate student was concerned over the current trend of AI automation decreasing jobs available for human developers:

\begin{quote}
    \textit{I wanted to be a developer, but there's news that big tech companies are letting go of developers and using AI instead. So that's why I feel stressed and worried.}
\end{quote}

Work experience was an additional factor that could compound or diminish one's anxiety over AI job displacement. Two undergraduate students expressed concern over being unable to secure an internship prior to graduation and entering a competitive job market without prior experience. One of them mentioned applying to as many openings as possible to maximize the possibility of gaining experience in any tech-related industry. On the contrary, one graduate student currently working in an AI-related position reported feeling more confident and less anxious in their abilities.

Two graduate students who intend to pursue AI also felt stressed that the field of AI itself may experience problems in the future. One student speculated that the push toward AI may lead to oversaturation and competition within the AI industry. Another student worried that even if they were able to obtain an AI-related position within a big tech company, they would still be at risk of being laid off.

\subsubsection{Ethical Concerns}

In addition to AI displacement of jobs, anxiety was expressed over ethical issues. Two students were concerned over the ethical implications of AI being used to develop itself in the future. One student mentioned that they fear AI collecting their personal data which could be used for ulterior motives such as mass surveillance. One student who was not worried about AI displacement of jobs mentioned that rather than job loss, their main concern was on AI's impacts on the environment, including water consumption, power consumption, and the placement of data centers used for AI development adversely impacting local communities. Finally, two students expressed interest in learning more about the ethics of AI at both the undergraduate and graduate degree levels within their courses.

\subsubsection{Optimism on Jobs}

Among two individuals who are relatively optimistic about their future career prospects, one student mentioned that AI was imperfect and made too many mistakes to be a serious contender for displacement of jobs by humans. Another student reasoned that the increased productivity in industry due to AI use may lead to short-term job losses, but could also encourage companies to hire more people to augment the productivity boost.

\subsection{AI Use}
\textit{Students are divided over whether AI should be used for ideation, with some wishing to avoid over-reliance while others prioritize efficiency.}

\subsubsection{Methods of AI Use}

Across interview participants, AI was used for general understanding, for tedious work, for ideation, to create practice problems, for programming, as an advanced search engine, to write emails, to improve language skills, to summarize academic papers, for writing, for feedback on ideas, for teaching, and to improve productivity in general. In particular, 12 students used AI for programming. Specific use cases included debugging, designing the front end of an application, checking over or improving already written code, and answering questions without needing to consult documentation.

\subsubsection{Ideation}

Participants were divided on whether AI should be used for ideation. Five students supported using AI for ideation while four students claimed that they do not. On the other hand, two students use AI to handle low-level work, giving them more time for high-level ideation.

Students who do not use AI for ideation cite avoiding overreliance as their motivation, as it might make them more replaceable. One student claimed that despite using it often, they avoid using it for entire tasks: 

\begin{quote}
    \textit{I use it a lot, kind of like an assistant, or like a thought partner. I won't really rely on it to do the entire job.}
\end{quote}

Moreover, one student claimed that AI only enables median competency at any task, but any further skill must come from themselves. One student stated that they try not to use AI if they cannot explain its output.

\subsubsection{Coursework}

All five first-year undergraduates claimed that AI can easily do programming tasks in first-year undergraduate courses. However, they all affirmed that first-year programming is still important for them to learn as it is foundational, and they asserted that AI should not be too heavily relied upon. One student claimed that AI was capable of writing first-year math proofs, while another student disagreed. Upper-year students claimed that AI could easily complete coursework for web development, software design, and computer systems. Meanwhile, they claimed that AI is less adept at advanced (graduate-level) or theory-focused courses.

\subsubsection{Reluctance to use AI}

Some students are reluctant to use AI but feel pressure to do so anyway. This is because they feel that their peers use AI and can write code more efficiently as a result, potentially making them more employable. However, one student claimed hesitancy to use AI because it removes the need to understand the output:

\begin{quote}
    \textit{I want to understand every line of code that I write. So I was hesitant to adopt AI because it kind of abstracts that away.}
\end{quote}

Furthermore, one student had a professor who recommended that they use AI. While this student tried to use AI for debugging code, they were only able to do so with mixed success. Five students expressed that AI can be misleading or incorrect, and another three expressed that AI is not useful without human guidance due to its lack of innate understanding.

\subsection{Vulnerable Fields to Displacement}
\textit{Students perceive software engineering, software development, and web development as replaceable by AI, whereas opinions are split on game development.}

\subsubsection{Software engineering and development}
13 of 25 students expressed that they had been discouraged from pursuing software engineering or software development roles. They noted the ability of AI to generate code efficiently, enabling the replacement of entry-level roles. Two students believed that AI was already better than a junior developer or software engineer, with one claiming the following:

\begin{quote}
    \textit{With AI's help, one senior developer can do three junior developers' jobs in much shorter time.}
\end{quote}

One student mentioned that despite having some background in software engineering, they found that AI still had more programming knowledge and were thus exploring other fields and positions. Another student expressed their anxiety that AI would replace software engineering roles, thereby pressuring students who would have wanted to become software engineers to pursue developer positions within the field of AI instead.

However, one student currently interning as a software engineer claimed that there are too many areas of software engineering for AI to replace all. Yet, they believe that AI can replace entry-level roles in the field, and they were discouraged from front-end roles. Another student who still expressed interest in pursuing software engineering felt pressured to learn AI tools for software engineering in order to appear more competitive to hiring companies.

\subsubsection{Front-end and web development}
7 students expressed that front-end work and web development are easily replaceable by AI. One student who uses AI to assist in front-end development claimed that ``anyone" can use AI to do it. With the perception of web development as easy to automate with AI, several undergraduate students felt discouraged from taking introductory courses in web development.

\subsubsection{Video game development}

5 students claimed that game development was replaceable while another 5 students believed it to be less replaceable than other roles. Students who view it to be replaceable cite that it is one of many programming-heavy tasks that can be automated and that AI is already used in the game development industry. Meanwhile, other students argued that game development has creative components such as art, music, storytelling, and design that are not easily done by AI. One student who was discouraged from game development still believes that design is less replaceable than programming.

\subsubsection{Miscellaneous}

Other roles students have been discouraged from include accounting, entry-level finance roles, DevOps, banking in general, and any ``repetitive work". One student claimed that management or HR could be replaced by AI despite those roles having a social component. Lastly, a first-year undergraduate student claimed that they have been discouraged from computer science as a whole:

\begin{quote}
    \textit{AI is replacing all these entry-level jobs. It's a lot of work, and at times I'm really skeptical, like, is this even worth it?}
\end{quote}

\subsection{Upskilling}
\textit{Students expressed many different methods of and rationales for upskilling. Industry trends have commonly led to an increased desire to use and learn about AI.}

\subsubsection{Pressure to Upskill}

With the rising popularity of Large Language Models (LLMs), students perceive that businesses want to increase AI use. Four students claimed that companies would use AI in lieu of employees since AI might be more efficient and decrease the human workers needed. One of them also mentioned that recent layoffs in big tech companies increased their anxiety. Many students feel pressured to become more efficient and keep up with current trends in industry by upskilling in AI. Several students believe that improving their AI proficiency will allow them to work in AI, which they perceive as a more secure role than software development or engineering. As a result, the increasing demand for AI by businesses compels students to gain AI-related skills to be employable.

One student expressed disinterest in AI as a field, yet decided to enroll in AI-related coursework to keep up with the industry:

\begin{quote}
    \textit{If it weren't for AI growing so much and changing the industry, I would probably not be doing the AI focus.}
\end{quote}

Another student mentioned that despite their dislike of studying AI and related algorithms, they feel pressured to study AI regardless. Another who was interested in robotics felt pressured to upskill in AI for the purposes of being able to secure a job in the robotics industry, in which they claim AI is becoming more prevalent. Overall, the increasing use of AI in businesses has also led to the perception that the number of available jobs for new graduates will decrease. Hence, upskilling is seen as necessary for future job security. This is consistent with \citet{wang_what_2022}'s result that AI job replacement anxiety positively affects extrinsic learning motivation.

\subsubsection{Graduate School}

Graduate school was commonly cited as a method for upskilling, although students' rationale for being motivated to seek graduate programs were varied. Several students who are currently enrolled in a professional computer science master's program expressed the necessity of upskilling to remain a competitive candidate for job and internship positions. Among undergraduates, graduate school is seen as both a means to gain more experience and AI knowledge and a fallback option in case a student cannot secure a job after graduating. Other students see graduate programs as the key to less replaceable roles or to opening more roles in the future. One student expressed the following:

\begin{quote}
    \textit{I didn't plan on going to graduate school when I enrolled, and now that's something I'm considering heavily.}
\end{quote}

\subsubsection{Additional Methods of Upskilling}

Other methods of upskilling were varied. Many students stated that they use AI tools to upskill. Some students interested in software engineering thought that knowing more AI technologies might help them stand out in a field seen as vulnerable to displacement by AI. Other students expressed interest in keeping pace with the current status of AI research and development or taking advantage of university programs and courses related to AI, such as adding a specialization in AI to their computer science degree or completing machine learning courses. In particular, among students who pursued a specialization in AI, one student mentioned that they felt pressured to because of AI but would not have done so if AI were less prevalent.

\subsubsection{Disinterest in Upskilling}

While the majority of students interviewed expressed interest in upskilling with AI, two individuals felt that they did not face additional pressure from AI to upskill. One student feels that since computer science has many subfields to specialize in, upskilling in AI is not a necessity as careers in other subfields would still exist. Another student does not feel threatened by the emergence of AI as AI still makes mistakes and does not have an inherent ``understanding" unlike humans.

\subsection{Reskilling}
\textit{As a way to become more secure against AI displacement of jobs, students have chosen to reskill by pursuing fields both within and outside of computer science perceived as more resilient and less likely to become automated.}

\subsubsection{Fields for reskilling}

Fields that students expressed interest in reskilling into were distributed across sciences, social sciences, engineering, and business. In case AI prevented them from securing jobs in computer science fields, some students chose to enroll in additional programs. Such programs include biology, physics, statistics, mathematics, and economics. One student considers social sciences as a backup because it requires a human component that is less replaceable. Another student finds that the human component of jobs like teachers and musicians cannot be replaced by AI.

\subsubsection{Specialized fields are less replaceable}

Within computer science subfields, one student switched their interests from software engineering to cybersecurity, which they perceived as safer from AI as they believe AI is worse at writing cybersecurity programs compared to software development. Areas of business mentioned as potential backup options included finance, marketing, and investing.

One student enrolled in a physics major to pursue quantum computing. They did not originally plan on pursuing it, believing upon entering university that any computer science degree brought security:

\begin{quote}
    \textit{When I enrolled in university for a computer science degree, I thought it's a pretty useful, pretty secure major. And now, I feel like you need to specialize in something to be seen as valuable.}
\end{quote}

However, a decrease in available entry-level roles and increase of AI use encouraged them to more seriously consider quantum computing since they believe that AI can not replace the physics component as easily as it can replace programming.

\subsubsection{Disinterest in Reskilling}

There were a few students who were not interested in reskilling. One student who was already motivated to study AI was not sure of what other skills to learn. Two students were uninterested in reskilling altogether, preferring to upskill with AI due to current trends. Another student did not want to reskill out of tech fields yet saw AI as dominant within those fields.

\subsection{International Student Pressure}
\textit{International students experience more pressure than domestic students since employment is necessary to secure permanent residence in Canada.}

\subsubsection{International Student Perspectives}

Six out of seven international students interviewed expressed that they faced pressure to get a job to secure permanent residence (PR). They felt that AI displacement of jobs put them at greater risk of having to leave Canada: 

\begin{quote}
    \textit{I think it puts more pressure on me because if I don't get a job soon enough, I might not be able to stay here.}
\end{quote}

One student expressed that they want PR so that they have access to opportunities both in Canada and their home country. One student expressed concern that companies prefer to hire domestic students over international students. Moreover, three students indicated a preference for joining the industry over graduate school to get permanent residence faster. Two of these students also cited that graduate school tuition is very expensive for international students. Lastly, one international student expressed lower anxiety due to already having secured employment in an AI-focused position, and claimed that \anon[their location of study]{Toronto} is an AI hub with many opportunities compared to their home country.

\subsubsection{Domestic Student Perspectives}

Of the 18 domestic students, 8 expressed that despite still being anxious they feel more secure in getting a job than if they were international students, with the primary reason that if they do not get a job they can remain in Canada. One domestic student claimed that companies may find it easier to hire domestic students. Another cited that domestic students do not have the pressure of renewing their VISA. A former international student who attained PR stated that they would not have applied to graduate school without having achieved permanent residence, due to the tuition fees.

Two domestic students claimed that international students may have more money, resources, or connections which may remove job opportunities from domestic students. Another student questioned why domestic status would matter, claiming that companies have quotas on the number of international students they must hire. Furthermore, they expressed that domestic students still need to be employed even if they are able to remain in Canada. 
\section{Discussion}

% Fairness: AI systems shouldn't have a differential effect on international and domestic students, but they do

% Accountability: How can we assign responsibility for AI job replacement anxiety? How can we mitigate it?

% Transparency: How can trust be built when designing systems in the future?

% Connect to deeper issues like migration, unemployment, education, etc.
    %%% Migration: universities in countries like Canada may face reductions in the number of international students enrolled if international students feel that it's too difficult to secure a job immediately after graduation
    %%% Unemployment: AI may exacerbate unemployment trends due to capitalist motives, ...?
    %%% Education: would education be devalued if AI can already do many things that undergrads can?
% Policymakers: 

\label{sec:discussion}

\subsection{Pressure on Job Security}

% It is a structural problem that the more specialized roles are locked behind further education and money. Often harder for international students. Problem since it's the specialized roles that students perceive to be less replaceable. Meanwhile entry level roles are easily replaceable

Our study found that more specialized roles are perceived as less replaceable by AI than entry-level roles such as junior development. An example of a specialized role is a quantum computing researcher, who may require not only education in computer science but also in physics \cite{juarez-ramirez_skills_2024}. However, less replaceable roles requiring more education pose a challenge to many students. Several students interviewed expressed that they previously believed that an undergraduate degree in computer science alone was sufficient since computer science seemed more secure in the past, but due to AI automation, they no longer feel this way. Hence, students are compelled to invest more time and resources into education to stay ahead. Students also expressed that entry-level roles becoming more accessible would decrease their job replacement anxiety, but AI automation makes this unlikely.

This is especially a problem for international students, who have to pay a much higher tuition for both undergraduate and graduate school. International students make up 32.4\% of all Canadian college and university enrollments \cite{statistics_canada_canadian_2025}. However, they may struggle to find a job in a more difficult market or lose their chance of obtaining permanent residence. Since the jobs attainable from their undergraduate degree are perceived as less secure, they may be compelled to pursue graduate school which may not be financially viable. International students also receive less funding opportunities than domestic students; for example, the Natural Sciences and Engineering Research Council of Canada's ``Undergraduate Student Research Awards" are only available to domestic students \cite{NSERC_USRA_2025}.

Historically, immigrants have seen Canada as a place of opportunity, with life satisfaction similar to that of the Canadian-born population \cite{monteiro_life_2022}. Moreover, Statistics Canada predicts that over half of the immigrant population in every major province feels a strong sense of belonging \cite{statistics_canada_immigrants_2023}. However, immigrants' life satisfaction in Canada is correlated with income, or economic security \cite{monteiro_life_2022}. Therefore, AI job replacement pressures make immigration more difficult for international students who wish to pursue a career in computer science. As computer science undergraduate enrollments have surged in the last few decades \cite{cra2017generationcs}, discouragement due to AI automation may not only decrease diversity in computer science workforces but also in the Canadian population itself. This is also applicable to other Western countries, such as the United States. Multiculturalism has been a part of Canada's identity for decades, with the country having one of the highest immigration rates in the world \cite{may_canada_2022}. The differential effect of AI job replacement on international students, discouraging them from pursuing computer science, runs counter to this identity and to the belief that individuals deserve equal opportunity regardless of where they come from. To ensure that international students are provided equitable support, entry-level roles attainable out of an undergraduate degree must not be displaced. 
% Therefore, AI automation has an unequal effect on domestic and international students, illustrating it as an unfair practice. 

Lastly, while our study found that domestic students feel more secure in their living arrangements, they feel similar anxiety about attaining entry-level computer science jobs. While graduate school tuition may be less expensive for domestic students, attaining a Master's degree or PhD is not only still costly but also a significant time commitment. Ultimately, it is in the best interests of policymakers to listen to impacted students, as they are at the forefront of this issue \cite{barnett_envisioning_2025}. Legislators can introduce tax benefits and financial incentives for corporations that retain human employees and foster partnerships between universities and industries to provide meaningful skill development for students \cite{qin_artificial_2024,jung_scoping_2024}.

% Ultimately, this calls for legislators to restrict businesses from using AI to automate the roles of junior developers. 

% Moreover, both academic institutions and governments should provide greater security for international students who wish to upskill by pursuing graduate school through funding awards.

\subsection{Industry Trends}

% People are discouraged from doing soft eng/dev, web dev
% Then these roles might become heavily automated, which could lead to a decline in the quality of our software, as AI still makes mistakes
% People will not understand how the code works in software, makes it harder to debug
% A lot of people will instead pursue specialized or research roles, this could have benefits
% Issue: A lot of people will try to join AI, it will become oversaturated

We found that over half of the interviewed students have been discouraged from software engineering, software development, front-end work, and web development. Furthermore, since students are upskilling by taking more AI courses, they will likely take less courses in software design. This may lead to a smaller number of graduates with the relevant skills, further motivating companies to automate those roles. However, a negative consequence of this could be a decline in the quality of software. According to one interview study, developers find it difficult to evaluate the trustworthiness of existing AI code generation tools \cite{wang_investigating_2024}. This is because LLMs are still prone to hallucination and often require multiple prompts to get a working response \cite{wang_investigating_2024, zhang_llm_2025}. Furthermore, if developers are more reliant on AI to write code, they will have a lesser understanding of how the code works. This would not only decrease trust in developers, but also make their roles even more automatable in the future. Lastly, this trend would make it difficult to assign accountability to developers for programming mistakes. Human developers follow principles such as ``Clean Architecture" \cite{martin_clean_2017} to avoid dependencies between different layers of a software. However, LLMs do not necessarily follow such a structure. This may make code more difficult to debug and create complex issues that are difficult to fix.

Another consequence of AI upskilling is that more students will work in AI. This is because students feel that AI requires more specialized knowledge to work in, as it is a research-heavy field. However, this may cause AI work to become over-saturated and job opportunities within the field to be more difficult to obtain. While companies will likely hire more workers in AI for the foreseeable future, the field may eventually slow down. One student expressed concern that AI may be a ``bubble", meaning that the escalation of AI profits greatly exceeds its intrinsic value, leaving the industry vulnerable to future losses. Previous research agrees with this sentiment, claiming that market excitement is outpacing the actual development of AI technology \cite{floridi_why_2024}. This may increase student anxiety since students who feel pressure to study AI may be unsure if it will be viable in the long term. Furthermore, it will worsen feelings of low job security for students pursuing AI in the workforce. This begs the question as to whether students should upskill by pursuing AI in the first place or simply pick a different subfield.

% Finally, the loss of workers in software engineering and development to more specialized roles may lead to an increase in research-oriented positions. This could have positive implications for research in industry, driving more rapid innovation. - commenting for now, might be a weaker argument

\subsection{Pressure to Use AI}

% Positive effects of upskilling

Through our study, we have verified that students feel pressure to upskill by learning how to use AI tools more effectively and by receiving an AI-related education. Moreover, even students who prefer not to use AI feel under pressure to use it to become more efficient and meet industry demands. In a previous study, AI learning anxiety is classified as debilitating, negatively affecting intrinsic and extrinsic learning motivation \cite{wang_what_2022}. However, our study suggests that AI learning anxiety can also be facilitating, since students believe that improving their AI proficiency will make them more employable, and allow them to work in AI. This suggests that it may increase extrinsic learning motivation.

The World Economic Forum's ``Future of Jobs Report" in 2025 surveyed employers for their planned strategies in response to AI's increasing prevalence \cite{forum_future_2025}. They found that 77\% of employers wish to reskill and upskill the existing workforce to work better alongside AI, 69\% wish to hire people with skills to design AI tools, 62\% wish to hire people with skills to better work alongside AI, and 41\% wish to downsize their workforce where AI can replicate people's work. Thus, in this context, pressure to use AI can encourage students to meet employers' standards, potentially increasing their employability. However, this is only the case because the alternative is unemployment, potentially making anxiety a bigger motivating factor than interest in the position.

% Negative effects of AI use for ideation

Pressure to use AI has even more negative effects. Our study illustrated that students are split on the use of AI for ideation, with roughly half for it and half against it. In work settings, efficiency and high output are prioritized due to capitalistic pressure \cite{sestino_leveraging_2022,fox_artificial_2024}. Thus, we argue that industrial and academic pressure will cause more students to use AI for ideation as a method to increase their productivity. These pressures may have detrimental effects on the cognition of employees in the long-term. One study found that LLMs are capable of generating more novel research ideas than human experts in the field of Natural Language Processing (NLP) \cite{si_can_2024}. The authors argue that while this may speed up research efficiency, overreliance on AI could lead to a decline in original human thought and reduce opportunities for human collaboration, which is often useful for refining ideas.

% What can be done

Since AI use has both positive and negative effects, educators should emphasize effective and ineffective ways to use AI. To determine these in greater certainty, further research should be done on whether varying types of AI use have a substantial effect on student career prospects and well being. For example, one could explore whether AI use for ideation and AI use for programming differentially affect academic performance, job placement, and self-efficacy.
\section{Limitations}
\label{sec:limitations}

This interview study was conducted at a single computer science department within one Canadian university. However, this university has a large and diverse student population represented in many different industries. Thus, we believe our findings are still representative of an average computer science postsecondary student population.

Since we performed snowball sampling, the resulting sample was dependent on the initial seed of participants. However, since the initial seed included various groups, the final distribution of participants across level of study, cumulative GPA, subfield of interest, and demographic characteristics was relatively equal. Therefore, we believe that our results effectively represent the sentiments of North American computer science students, and that they are not heavily affected by biases.

Although we received valuable international student perspectives, we were only able to interview 7 of them, while 18 domestic students were interviewed. This was a result of the snowball sampling seed. The strong agreement between the answers of the international students to Question 8 of the Interview Protocol (\ref{methods:protocol}) ensures the credibility of our findings. However, future qualitative or quantitative studies on AI job replacement anxiety in international computer science students may discover perspectives that our study missed.

Lastly, we acknowledge the presence of selection bias in our study, namely that students who are anxious about AI job displacement are more likely to join the study than students who are not. However, the incentive of \$10 CAD Amazon gift cards would appeal to both students who do and do not experience AI job replacement anxiety.

Future research may address these limitations by conducting interviews with students across several universities in several countries. Furthermore, stratified sampling could be performed to get an equal amount of domestic and international student perspectives, and better analyze different characteristics such as other demographic information or level of study.

\section{Conclusion}

\label{sec:conclusions}

Computer science students entering the workforce face significant pressure in the AI era. Through our study, we observed that students are worried that businesses will use AI to automate programming, decreasing their opportunities for employment. This discourages students from entering subfields such as software engineering and web development. Meanwhile, specialized fields such as AI research and quantum computing are seen as more secure from automation and displacement by AI. To cope with this pressure, most students upskill. Academically, students feel pressure to enroll in AI coursework or pursue graduate school. Industrially, students learn how to use AI tools to increase employability. Students also reskill to cope with AI job replacement anxiety. Outside of computer science, students choose to study fields from many areas including biology, physics, economics, and business in hopes of building a more diverse skill set. Moreover, international students face additional pressure to find a job and secure permanent residence, causing AI automation of entry-level jobs to be particularly damaging for them.

Job displacement by AI is an emerging issue that has implications for the livelihoods of many students and professionals, beyond the computer science student community at the university where the study was conducted. While corporations view automation of work as advantageous for increasing efficiency while decreasing costs, these systems cause employees to feel insecurity when navigating the industry and job market. Furthermore, while the field of computer science experienced significant growth over the last few decades, job replacement anxiety may discourage students from pursuing it if AI job displacement trends continue.

% Students use AI to varying degrees based on their perception of AI. While some students use AI almost daily to assist with tasks from programming to ideation, others use it less frequently or try to avoid it altogether. Some students noted that despite the hype behind AI and business adoption of AI, it is far from perfect and prone to making errors. Others pointed out that AI could not truly replicate human innovation, interpretation, and feeling. Ethical issues students are concerned about include the use of AI to develop itself, privacy and surveillance, and environmental impacts.

\section*{Generative AI Usage Statement}
No generative AI was used in the process of this study or in the writing of this paper.

\newpage

\bibliographystyle{ACM-Reference-Format}
\bibliography{references}


\begin{thebibliography}{71}

%%% the Web address.

\ifx \showCODEN    \undefined \def \showCODEN     #1{\unskip}     \fi
\ifx \showISBNx    \undefined \def \showISBNx     #1{\unskip}     \fi
\ifx \showISBNxiii \undefined \def \showISBNxiii  #1{\unskip}     \fi
\ifx \showISSN     \undefined \def \showISSN      #1{\unskip}     \fi
\ifx \showLCCN     \undefined \def \showLCCN      #1{\unskip}     \fi
\ifx \shownote     \undefined \def \shownote      #1{#1}          \fi
\ifx \showarticletitle \undefined \def \showarticletitle #1{#1}   \fi
\ifx \showURL      \undefined \def \showURL       {\relax}        \fi
% invisible to TeX
\providecommand\bibfield[2]{#2}
\providecommand\bibinfo[2]{#2}
\providecommand\natexlab[1]{#1}
\providecommand\showeprint[2][]{arXiv:#2}

\bibitem[Acemoglu and Restrepo(2020)]%
        {acemoglu_wrong_2020}
\bibfield{author}{\bibinfo{person}{Daron Acemoglu} {and} \bibinfo{person}{Pascual Restrepo}.} \bibinfo{year}{2020}\natexlab{}.
\newblock \showarticletitle{The wrong kind of {AI}? {Artificial} intelligence and the future of labour demand}.
\newblock \bibinfo{journal}{\emph{Cambridge Journal of Regions, Economy and Society}} \bibinfo{volume}{13}, \bibinfo{number}{1} (\bibinfo{date}{May} \bibinfo{year}{2020}), \bibinfo{pages}{25--35}.
\newblock
\showISSN{1752-1378, 1752-1386}
\href{https://doi.org/10.1093/cjres/rsz022}{doi:\nolinkurl{10.1093/cjres/rsz022}}


\bibitem[Ahmed et~al\mbox{.}(2025)]%
        {ahmed_artificial_2025}
\bibfield{author}{\bibinfo{person}{Iftekhar Ahmed}, \bibinfo{person}{Aldeida Aleti}, \bibinfo{person}{Haipeng Cai}, \bibinfo{person}{Alexander Chatzigeorgiou}, \bibinfo{person}{Pinjia He}, \bibinfo{person}{Xing Hu}, \bibinfo{person}{Mauro Pezzè}, \bibinfo{person}{Denys Poshyvanyk}, {and} \bibinfo{person}{Xin Xia}.} \bibinfo{year}{2025}\natexlab{}.
\newblock \showarticletitle{Artificial {Intelligence} for {Software} {Engineering}: {The} {Journey} {So} {Far} and the {Road} {Ahead}}.
\newblock \bibinfo{journal}{\emph{ACM Transactions on Software Engineering and Methodology}} \bibinfo{volume}{34}, \bibinfo{number}{5} (\bibinfo{date}{June} \bibinfo{year}{2025}), \bibinfo{pages}{1--27}.
\newblock
\showISSN{1049-331X, 1557-7392}
\href{https://doi.org/10.1145/3719006}{doi:\nolinkurl{10.1145/3719006}}


\bibitem[Barnett et~al\mbox{.}(2025)]%
        {barnett_envisioning_2025}
\bibfield{author}{\bibinfo{person}{Julia Barnett}, \bibinfo{person}{Kimon Kieslich}, \bibinfo{person}{Natali Helberger}, {and} \bibinfo{person}{Nicholas Diakopoulos}.} \bibinfo{year}{2025}\natexlab{}.
\newblock \showarticletitle{Envisioning {Stakeholder}-{Action} {Pairs} to {Mitigate} {Negative} {Impacts} of {AI}: {A} {Participatory} {Approach} to {Inform} {Policy} {Making}}. In \bibinfo{booktitle}{\emph{Proceedings of the 2025 {ACM} {Conference} on {Fairness}, {Accountability}, and {Transparency}}}. \bibinfo{publisher}{ACM}, \bibinfo{address}{Athens Greece}, \bibinfo{pages}{1424--1449}.
\newblock
\showISBNx{979-8-4007-1482-5}
\href{https://doi.org/10.1145/3715275.3732096}{doi:\nolinkurl{10.1145/3715275.3732096}}


\bibitem[Biernacki and Waldorf(1981)]%
        {biernacki_snowball_1981}
\bibfield{author}{\bibinfo{person}{Patrick Biernacki} {and} \bibinfo{person}{Dan Waldorf}.} \bibinfo{year}{1981}\natexlab{}.
\newblock \showarticletitle{Snowball {Sampling}: {Problems} and {Techniques} of {Chain} {Referral} {Sampling}}.
\newblock \bibinfo{journal}{\emph{Sociological Methods \& Research}} \bibinfo{volume}{10}, \bibinfo{number}{2} (\bibinfo{date}{Nov.} \bibinfo{year}{1981}), \bibinfo{pages}{141--163}.
\newblock
\showISSN{0049-1241, 1552-8294}
\href{https://doi.org/10.1177/004912418101000205}{doi:\nolinkurl{10.1177/004912418101000205}}


\bibitem[Chan and Hu(2023)]%
        {chan_students_2023}
\bibfield{author}{\bibinfo{person}{Cecilia Ka~Yuk Chan} {and} \bibinfo{person}{Wenjie Hu}.} \bibinfo{year}{2023}\natexlab{}.
\newblock \showarticletitle{Students’ voices on generative {AI}: perceptions, benefits, and challenges in higher education}.
\newblock \bibinfo{journal}{\emph{International Journal of Educational Technology in Higher Education}} \bibinfo{volume}{20}, \bibinfo{number}{1} (\bibinfo{date}{July} \bibinfo{year}{2023}), \bibinfo{pages}{43}.
\newblock
\showISSN{2365-9440}
\href{https://doi.org/10.1186/s41239-023-00411-8}{doi:\nolinkurl{10.1186/s41239-023-00411-8}}


\bibitem[Chung et~al\mbox{.}(2025)]%
        {chung_artificial_2025}
\bibfield{author}{\bibinfo{person}{Yang~Woon Chung}, \bibinfo{person}{Seunghee Im}, \bibinfo{person}{Jung~Eun Kim}, {and} \bibinfo{person}{Jeong~Kwon Yun}.} \bibinfo{year}{2025}\natexlab{}.
\newblock \showarticletitle{Artificial intelligence awareness, career resilience, job insecurity and behavioural outcomes}.
\newblock \bibinfo{journal}{\emph{Australian Journal of Psychology}} \bibinfo{volume}{77}, \bibinfo{number}{1} (\bibinfo{date}{Dec.} \bibinfo{year}{2025}), \bibinfo{pages}{2559910}.
\newblock
\showISSN{0004-9530, 1742-9536}
\href{https://doi.org/10.1080/00049530.2025.2559910}{doi:\nolinkurl{10.1080/00049530.2025.2559910}}


\bibitem[Clos and Chen(2024)]%
        {clos_investigating_2024}
\bibfield{author}{\bibinfo{person}{Jeremie Clos} {and} \bibinfo{person}{Yoke~Yie Chen}.} \bibinfo{year}{2024}\natexlab{}.
\newblock \showarticletitle{Investigating the {Impact} of {Generative} {AI} on {Students} and {Educators}: {Evidence} and {Insights} from the {Literature}}. In \bibinfo{booktitle}{\emph{Proceedings of the {Second} {International} {Symposium} on {Trustworthy} {Autonomous} {Systems}}}. \bibinfo{publisher}{ACM}, \bibinfo{address}{Austin TX USA}, \bibinfo{pages}{1--6}.
\newblock
\showISBNx{979-8-4007-0989-0}
\href{https://doi.org/10.1145/3686038.3686063}{doi:\nolinkurl{10.1145/3686038.3686063}}


\bibitem[Committee(2017)]%
        {cra2017generationcs}
\bibfield{author}{\bibinfo{person}{{Computing Research Association}~Enrollment Committee}.} \bibinfo{year}{2017}\natexlab{}.
\newblock \bibinfo{booktitle}{\emph{Generation CS: Computer Science Undergraduate Enrollments Surge Since 2006}}.
\newblock \bibinfo{type}{{T}echnical {R}eport}. \bibinfo{institution}{Computing Research Association}.
\newblock
\urldef\tempurl%
\url{https://cra.org/data/Generation-CS/}
\showURL{%
\tempurl}
\newblock
\shownote{Accessed: 2026-01-10}.


\bibitem[Constantinides and Quercia(2025)]%
        {constantinides_ai_2025}
\bibfield{author}{\bibinfo{person}{Marios Constantinides} {and} \bibinfo{person}{Daniele Quercia}.} \bibinfo{year}{2025}\natexlab{}.
\newblock \showarticletitle{{AI}, {Jobs}, and the {Automation} {Trap}: {Where} {Is} {HCI}?}. In \bibinfo{booktitle}{\emph{Proceedings of the 4th {Annual} {Symposium} on {Human}-{Computer} {Interaction} for {Work}}}. \bibinfo{publisher}{ACM}, \bibinfo{address}{Amsterdam Netherlands}, \bibinfo{pages}{1--8}.
\newblock
\showISBNx{979-8-4007-1384-2}
\href{https://doi.org/10.1145/3729176.3729191}{doi:\nolinkurl{10.1145/3729176.3729191}}


\bibitem[Cordasco and Véliz(2025)]%
        {cordasco_self-esteem_2025}
\bibfield{author}{\bibinfo{person}{Carlo~Ludovico Cordasco} {and} \bibinfo{person}{Carissa Véliz}.} \bibinfo{year}{2025}\natexlab{}.
\newblock \showarticletitle{Self-{Esteem} and {Technological} {Unemployment}: {Should} {We} {Halt} {AI} to {Protect} {Meaningful} {Work}?}
\newblock \bibinfo{journal}{\emph{Journal of Business Ethics}} \bibinfo{volume}{202}, \bibinfo{number}{1} (\bibinfo{date}{Nov.} \bibinfo{year}{2025}), \bibinfo{pages}{21--33}.
\newblock
\showISSN{0167-4544, 1573-0697}
\href{https://doi.org/10.1007/s10551-025-06007-8}{doi:\nolinkurl{10.1007/s10551-025-06007-8}}


\bibitem[Cramarenco et~al\mbox{.}(2023)]%
        {cramarenco_impact_2023}
\bibfield{author}{\bibinfo{person}{Romana~Emilia Cramarenco}, \bibinfo{person}{Monica~Ioana Burcă-Voicu}, {and} \bibinfo{person}{Dan~Cristian Dabija}.} \bibinfo{year}{2023}\natexlab{}.
\newblock \showarticletitle{The impact of artificial intelligence ({AI}) on employees’ skills and well-being in global labor markets: {A} systematic review}.
\newblock \bibinfo{journal}{\emph{Oeconomia Copernicana}} \bibinfo{volume}{14}, \bibinfo{number}{3} (\bibinfo{date}{Sept.} \bibinfo{year}{2023}), \bibinfo{pages}{731--767}.
\newblock
\showISSN{2353-1827, 2083-1277}
\href{https://doi.org/10.24136/oc.2023.022}{doi:\nolinkurl{10.24136/oc.2023.022}}


\bibitem[Dahlin(2024)]%
        {dahlin_who_2024}
\bibfield{author}{\bibinfo{person}{Eric Dahlin}.} \bibinfo{year}{2024}\natexlab{}.
\newblock \showarticletitle{Who {Says} {Artificial} {Intelligence} {Is} {Stealing} {Our} {Jobs}?}
\newblock \bibinfo{journal}{\emph{Socius: Sociological Research for a Dynamic World}}  \bibinfo{volume}{10} (\bibinfo{date}{Jan.} \bibinfo{year}{2024}), \bibinfo{pages}{23780231241259672}.
\newblock
\showISSN{2378-0231, 2378-0231}
\href{https://doi.org/10.1177/23780231241259672}{doi:\nolinkurl{10.1177/23780231241259672}}


\bibitem[Denning(2025)]%
        {denning_artificial_2025}
\bibfield{author}{\bibinfo{person}{Peter~J. Denning}.} \bibinfo{year}{2025}\natexlab{}.
\newblock \showarticletitle{Artificial {Intelligence}: {Generative} {AI}}.
\newblock \bibinfo{journal}{\emph{Ubiquity}} \bibinfo{volume}{2025}, \bibinfo{number}{July} (\bibinfo{date}{April} \bibinfo{year}{2025}), \bibinfo{pages}{1--19}.
\newblock
\showISSN{1530-2180}
\href{https://doi.org/10.1145/3747356}{doi:\nolinkurl{10.1145/3747356}}


\bibitem[Denning and Denning(2020)]%
        {denning_dilemmas_2020}
\bibfield{author}{\bibinfo{person}{Peter~J. Denning} {and} \bibinfo{person}{Dorothy~E. Denning}.} \bibinfo{year}{2020}\natexlab{}.
\newblock \showarticletitle{Dilemmas of artificial intelligence}.
\newblock \bibinfo{journal}{\emph{Commun. ACM}} \bibinfo{volume}{63}, \bibinfo{number}{3} (\bibinfo{date}{Feb.} \bibinfo{year}{2020}), \bibinfo{pages}{22--24}.
\newblock
\showISSN{0001-0782, 1557-7317}
\href{https://doi.org/10.1145/3379920}{doi:\nolinkurl{10.1145/3379920}}


\bibitem[Dou(2023)]%
        {dou_does_2023}
\bibfield{author}{\bibinfo{person}{Yixuan Dou}.} \bibinfo{year}{2023}\natexlab{}.
\newblock \showarticletitle{Does the application of artificial intelligence technology affect employees' turnover intention?}. In \bibinfo{booktitle}{\emph{Proceedings of the 2023 6th {International} {Conference} on {Information} {Management} and {Management} {Science}}}. \bibinfo{publisher}{ACM}, \bibinfo{address}{Chengdu China}, \bibinfo{pages}{283--287}.
\newblock
\showISBNx{979-8-4007-0768-1}
\href{https://doi.org/10.1145/3625469.3625488}{doi:\nolinkurl{10.1145/3625469.3625488}}


\bibitem[Floridi(2024)]%
        {floridi_why_2024}
\bibfield{author}{\bibinfo{person}{Luciano Floridi}.} \bibinfo{year}{2024}\natexlab{}.
\newblock \showarticletitle{Why the {AI} {Hype} is {Another} {Tech} {Bubble}}.
\newblock \bibinfo{journal}{\emph{Philosophy \& Technology}} \bibinfo{volume}{37}, \bibinfo{number}{4} (\bibinfo{date}{Dec.} \bibinfo{year}{2024}), \bibinfo{pages}{128, s13347--024--00817--w}.
\newblock
\showISSN{2210-5433, 2210-5441}
\href{https://doi.org/10.1007/s13347-024-00817-w}{doi:\nolinkurl{10.1007/s13347-024-00817-w}}


\bibitem[Forum(2025)]%
        {forum_future_2025}
\bibfield{author}{\bibinfo{person}{World~Economic Forum}.} \bibinfo{year}{2025}\natexlab{}.
\newblock \bibinfo{booktitle}{\emph{The {Future} of {Jobs} {Report} 2025}}.
\newblock \bibinfo{type}{{T}echnical {R}eport}. \bibinfo{institution}{World Economic Forum}.
\newblock
\urldef\tempurl%
\url{https://www.weforum.org/publications/the-future-of-jobs-report-2025/}
\showURL{%
\tempurl}


\bibitem[Fox(2024)]%
        {fox_artificial_2024}
\bibfield{author}{\bibinfo{person}{Nick~J. Fox}.} \bibinfo{year}{2024}\natexlab{}.
\newblock \showarticletitle{Artificial {Intelligence} and the {Black} {Hole} of {Capitalism}: {A} {More}-than-{Human} {Political} {Ethology}}.
\newblock \bibinfo{journal}{\emph{Social Sciences}} \bibinfo{volume}{13}, \bibinfo{number}{10} (\bibinfo{date}{Sept.} \bibinfo{year}{2024}), \bibinfo{pages}{507}.
\newblock
\showISSN{2076-0760}
\href{https://doi.org/10.3390/socsci13100507}{doi:\nolinkurl{10.3390/socsci13100507}}


\bibitem[Frank et~al\mbox{.}(2025)]%
        {frank_ai_2025}
\bibfield{author}{\bibinfo{person}{Morgan~R Frank}, \bibinfo{person}{Yong-Yeol Ahn}, {and} \bibinfo{person}{Esteban Moro}.} \bibinfo{year}{2025}\natexlab{}.
\newblock \showarticletitle{{AI} exposure predicts unemployment risk: {A} new approach to technology-driven job loss}.
\newblock \bibinfo{journal}{\emph{PNAS Nexus}} \bibinfo{volume}{4}, \bibinfo{number}{4} (\bibinfo{date}{March} \bibinfo{year}{2025}), \bibinfo{pages}{pgaf107}.
\newblock
\showISSN{2752-6542}
\href{https://doi.org/10.1093/pnasnexus/pgaf107}{doi:\nolinkurl{10.1093/pnasnexus/pgaf107}}


\bibitem[Gambacorta et~al\mbox{.}(2024)]%
        {gambacorta_generative_2024}
\bibfield{author}{\bibinfo{person}{Leonardo Gambacorta}, \bibinfo{person}{Han Qiu}, \bibinfo{person}{Sharon Shan}, {and} \bibinfo{person}{Daniel Rees}.} \bibinfo{year}{2024}\natexlab{}.
\newblock \bibinfo{booktitle}{\emph{Generative {AI} and labour productivity: a field experiment on coding}}.
\newblock \bibinfo{type}{{BIS} {Working} {Papers}} 1208. \bibinfo{institution}{Bank for International Settlements}.
\newblock
\urldef\tempurl%
\url{https://www.bis.org/publ/work1208.htm}
\showURL{%
\tempurl}


\bibitem[George et~al\mbox{.}(2023)]%
        {george_chatgpt_2023}
\bibfield{author}{\bibinfo{person}{Dr.A.Shaji George}, \bibinfo{person}{{A.S.Hovan George}}, {and} \bibinfo{person}{{A.S.Gabrio Martin}}.} \bibinfo{year}{2023}\natexlab{}.
\newblock \showarticletitle{{ChatGPT} and the {Future} of {Work}: {A} {Comprehensive} {Analysis} of {AI}'s {Impact} on {Jobs} and {Employment}}.
\newblock  (\bibinfo{date}{June} \bibinfo{year}{2023}).
\newblock
\href{https://doi.org/10.5281/ZENODO.8076921}{doi:\nolinkurl{10.5281/ZENODO.8076921}}
\newblock
\shownote{Publisher: Zenodo}.


\bibitem[Ghotbi et~al\mbox{.}(2022)]%
        {ghotbi_attitude_2022}
\bibfield{author}{\bibinfo{person}{Nader Ghotbi}, \bibinfo{person}{Manh~Tung Ho}, {and} \bibinfo{person}{Peter Mantello}.} \bibinfo{year}{2022}\natexlab{}.
\newblock \showarticletitle{Attitude of college students towards ethical issues of artificial intelligence in an international university in {Japan}}.
\newblock \bibinfo{journal}{\emph{AI \& SOCIETY}} \bibinfo{volume}{37}, \bibinfo{number}{1} (\bibinfo{date}{March} \bibinfo{year}{2022}), \bibinfo{pages}{283--290}.
\newblock
\showISSN{0951-5666, 1435-5655}
\href{https://doi.org/10.1007/s00146-021-01168-2}{doi:\nolinkurl{10.1007/s00146-021-01168-2}}


\bibitem[Golda et~al\mbox{.}(2024)]%
        {golda_privacy_2024}
\bibfield{author}{\bibinfo{person}{Abenezer Golda}, \bibinfo{person}{Kidus Mekonen}, \bibinfo{person}{Amit Pandey}, \bibinfo{person}{Anushka Singh}, \bibinfo{person}{Vikas Hassija}, \bibinfo{person}{Vinay Chamola}, {and} \bibinfo{person}{Biplab Sikdar}.} \bibinfo{year}{2024}\natexlab{}.
\newblock \showarticletitle{Privacy and {Security} {Concerns} in {Generative} {AI}: {A} {Comprehensive} {Survey}}.
\newblock \bibinfo{journal}{\emph{IEEE Access}}  \bibinfo{volume}{12} (\bibinfo{year}{2024}), \bibinfo{pages}{48126--48144}.
\newblock
\showISSN{2169-3536}
\href{https://doi.org/10.1109/ACCESS.2024.3381611}{doi:\nolinkurl{10.1109/ACCESS.2024.3381611}}


\bibitem[Goodman(1961)]%
        {goodman_snowball_1961}
\bibfield{author}{\bibinfo{person}{Leo~A. Goodman}.} \bibinfo{year}{1961}\natexlab{}.
\newblock \showarticletitle{Snowball {Sampling}}.
\newblock \bibinfo{journal}{\emph{The Annals of Mathematical Statistics}} \bibinfo{volume}{32}, \bibinfo{number}{1} (\bibinfo{date}{March} \bibinfo{year}{1961}), \bibinfo{pages}{148--170}.
\newblock
\showISSN{0003-4851}
\href{https://doi.org/10.1214/aoms/1177705148}{doi:\nolinkurl{10.1214/aoms/1177705148}}


\bibitem[Ji et~al\mbox{.}(2024)]%
        {ji_current_2024}
\bibfield{author}{\bibinfo{person}{Wei Ji}, \bibinfo{person}{Jian Sun}, \bibinfo{person}{Biaoxin Chen}, {and} \bibinfo{person}{Chuangli Luo}.} \bibinfo{year}{2024}\natexlab{}.
\newblock \showarticletitle{The {Current} {Status} and {Trends} of {Research} on the {Impact} of {Generative} {Artificial} {Intelligence} on {Employment} in {China}}. In \bibinfo{booktitle}{\emph{Proceedings of the 2024 {Guangdong}-{Hong} {Kong}-{Macao} {Greater} {Bay} {Area} {International} {Conference} on {Digital} {Economy} and {Artificial} {Intelligence}}}. \bibinfo{publisher}{ACM}, \bibinfo{address}{Hongkong China}, \bibinfo{pages}{948--953}.
\newblock
\showISBNx{979-8-4007-1714-7}
\href{https://doi.org/10.1145/3675417.3675574}{doi:\nolinkurl{10.1145/3675417.3675574}}


\bibitem[Jung et~al\mbox{.}(2024)]%
        {jung_scoping_2024}
\bibfield{author}{\bibinfo{person}{Jisun Jung}, \bibinfo{person}{Yutong Wang}, {and} \bibinfo{person}{Mabel Sanchez~Barrioluengo}.} \bibinfo{year}{2024}\natexlab{}.
\newblock \showarticletitle{A scoping review on graduate employability in an era of ‘{Technological} {Unemployment}’}.
\newblock \bibinfo{journal}{\emph{Higher Education Research \& Development}} \bibinfo{volume}{43}, \bibinfo{number}{3} (\bibinfo{date}{April} \bibinfo{year}{2024}), \bibinfo{pages}{542--562}.
\newblock
\showISSN{0729-4360, 1469-8366}
\href{https://doi.org/10.1080/07294360.2023.2292660}{doi:\nolinkurl{10.1080/07294360.2023.2292660}}


\bibitem[Juárez-Ramírez et~al\mbox{.}(2024)]%
        {juarez-ramirez_skills_2024}
\bibfield{author}{\bibinfo{person}{Reyes Juárez-Ramírez}, \bibinfo{person}{Samantha Jiménez}, \bibinfo{person}{Christian~X. Navarro}, \bibinfo{person}{César Guerra-García}, \bibinfo{person}{Hector~G. Perez-Gonzalez}, \bibinfo{person}{Carlos Fernández-y Fernández}, \bibinfo{person}{Javier Ortiz-Hernández}, {and} \bibinfo{person}{Karina Cancino}.} \bibinfo{year}{2024}\natexlab{}.
\newblock \showarticletitle{Skills {Required} for {Quantum} {Computing}: {A} {Comprehensive} {Review} of {Recent} {Studies}}.
\newblock \bibinfo{journal}{\emph{Programming and Computer Software}} \bibinfo{volume}{50}, \bibinfo{number}{8} (\bibinfo{date}{Dec.} \bibinfo{year}{2024}), \bibinfo{pages}{844--874}.
\newblock
\showISSN{0361-7688, 1608-3261}
\href{https://doi.org/10.1134/S0361768824700804}{doi:\nolinkurl{10.1134/S0361768824700804}}


\bibitem[Kam et~al\mbox{.}(2025)]%
        {kam_what_2025}
\bibfield{author}{\bibinfo{person}{Matthew Kam}, \bibinfo{person}{Cody Miller}, \bibinfo{person}{Miaoxin Wang}, \bibinfo{person}{Abey Tidwell}, \bibinfo{person}{Irene~A. Lee}, \bibinfo{person}{Joyce Malyn-Smith}, \bibinfo{person}{Beatriz Perret}, \bibinfo{person}{Vikram Tiwari}, \bibinfo{person}{Joshua Kenitzer}, \bibinfo{person}{Andrew Macvean}, {and} \bibinfo{person}{Erin Barrar}.} \bibinfo{year}{2025}\natexlab{}.
\newblock \showarticletitle{What do professional software developers need to know to succeed in an age of {Artificial} {Intelligence}?}. In \bibinfo{booktitle}{\emph{Proceedings of the 33rd {ACM} {International} {Conference} on the {Foundations} of {Software} {Engineering}}}. \bibinfo{publisher}{ACM}, \bibinfo{address}{Clarion Hotel Trondheim Trondheim Norway}, \bibinfo{pages}{947--958}.
\newblock
\showISBNx{979-8-4007-1276-0}
\href{https://doi.org/10.1145/3696630.3727251}{doi:\nolinkurl{10.1145/3696630.3727251}}


\bibitem[Kudoh and Miyamoto(2025)]%
        {kudoh_robots_2025}
\bibfield{author}{\bibinfo{person}{Noritaka Kudoh} {and} \bibinfo{person}{Hiroaki Miyamoto}.} \bibinfo{year}{2025}\natexlab{}.
\newblock \showarticletitle{Robots, {AI}, and unemployment}.
\newblock \bibinfo{journal}{\emph{Journal of Economic Dynamics and Control}}  \bibinfo{volume}{174} (\bibinfo{date}{May} \bibinfo{year}{2025}), \bibinfo{pages}{105069}.
\newblock
\showISSN{01651889}
\href{https://doi.org/10.1016/j.jedc.2025.105069}{doi:\nolinkurl{10.1016/j.jedc.2025.105069}}


\bibitem[Li(2024)]%
        {li_reskilling_2024}
\bibfield{author}{\bibinfo{person}{Ling Li}.} \bibinfo{year}{2024}\natexlab{}.
\newblock \showarticletitle{Reskilling and {Upskilling} the {Future}-ready {Workforce} for {Industry} 4.0 and {Beyond}}.
\newblock \bibinfo{journal}{\emph{Information Systems Frontiers}} \bibinfo{volume}{26}, \bibinfo{number}{5} (\bibinfo{date}{Oct.} \bibinfo{year}{2024}), \bibinfo{pages}{1697--1712}.
\newblock
\showISSN{1387-3326, 1572-9419}
\href{https://doi.org/10.1007/s10796-022-10308-y}{doi:\nolinkurl{10.1007/s10796-022-10308-y}}


\bibitem[Liu(2025)]%
        {liu_what_2025}
\bibfield{author}{\bibinfo{person}{Larry Liu}.} \bibinfo{year}{2025}\natexlab{}.
\newblock \showarticletitle{What is the {Future} of {Work} in the {Generative} {AI} {Era}? {A} {Marxist} and {Ricardian} {Analysis}}.
\newblock \bibinfo{journal}{\emph{tripleC: Communication, Capitalism \& Critique. Open Access Journal for a Global Sustainable Information Society}} \bibinfo{volume}{23}, \bibinfo{number}{1} (\bibinfo{date}{March} \bibinfo{year}{2025}), \bibinfo{pages}{131--148}.
\newblock
\showISSN{1726-670X, 1726-670X}
\href{https://doi.org/10.31269/triplec.v23i1.1536}{doi:\nolinkurl{10.31269/triplec.v23i1.1536}}


\bibitem[Malheiros et~al\mbox{.}(2024)]%
        {malheiros_impact_2024}
\bibfield{author}{\bibinfo{person}{Phelipe~Silva Malheiros}, \bibinfo{person}{Rayfran~Rocha Lima}, {and} \bibinfo{person}{Ana~Carolina Oran}.} \bibinfo{year}{2024}\natexlab{}.
\newblock \showarticletitle{Impact of {Generative} {AI} {Technologies} on {Software} {Development} {Professionals}' {Perceptions} of {Job} {Security}}. In \bibinfo{booktitle}{\emph{Proceedings of the {XXIII} {Brazilian} {Symposium} on {Software} {Quality}}}. \bibinfo{publisher}{ACM}, \bibinfo{address}{Salvador Bahia Brazil}, \bibinfo{pages}{169--178}.
\newblock
\showISBNx{979-8-4007-1777-2}
\href{https://doi.org/10.1145/3701625.3701656}{doi:\nolinkurl{10.1145/3701625.3701656}}


\bibitem[Manyika et~al\mbox{.}(2017)]%
        {manyika_jobs_2017}
\bibfield{author}{\bibinfo{person}{James Manyika}, \bibinfo{person}{Susan Lund}, \bibinfo{person}{Jacques Bughin}, \bibinfo{person}{Jonathan Woetzel}, \bibinfo{person}{Parul Batra}, \bibinfo{person}{Saurabh Sanghvi}, \bibinfo{person}{Ryan Ko}, {and} \bibinfo{person}{Michael Chui}.} \bibinfo{year}{2017}\natexlab{}.
\newblock \bibinfo{booktitle}{\emph{Jobs {Lost}, {Jobs} {Gained}: {Workforce} {Transitions} in a {Time} of {Automation}}}.
\newblock \bibinfo{type}{{MGI} {Report}} December 2017. \bibinfo{institution}{McKinsey Global Institute}, \bibinfo{address}{San Francisco, CA}. \bibinfo{pages}{148} pages.
\newblock
\urldef\tempurl%
\url{https://www.mckinsey.com/~/media/mckinsey/industries/public%20and%20social%20sector/our%20insights/what%20the%20future%20of%20work%20will%20mean%20for%20jobs%20skills%20and%20wages/mgi%20jobs%20lost-jobs%20gained_report_december%202017.pdf}
\showURL{%
\tempurl}


\bibitem[Martin(2017)]%
        {martin_clean_2017}
\bibfield{author}{\bibinfo{person}{Robert~C. Martin}.} \bibinfo{year}{2017}\natexlab{}.
\newblock \bibinfo{booktitle}{\emph{Clean Architecture: A Craftsman's Guide to Software Structure and Design} (\bibinfo{edition}{1st} ed.)}.
\newblock \bibinfo{publisher}{Prentice Hall Press}, \bibinfo{address}{USA}.
\newblock
\showISBNx{0134494164}


\bibitem[May(2022)]%
        {may_canada_2022}
\bibfield{author}{\bibinfo{person}{Paul May}.} \bibinfo{year}{2022}\natexlab{}.
\newblock \showarticletitle{Canada: the standard bearer of multiculturalism in the world? {An} analysis of the {Canadian} public debate on multiculturalism (2010–2020)}.
\newblock \bibinfo{journal}{\emph{Ethnic and Racial Studies}} \bibinfo{volume}{45}, \bibinfo{number}{10} (\bibinfo{date}{July} \bibinfo{year}{2022}), \bibinfo{pages}{1939--1960}.
\newblock
\showISSN{0141-9870, 1466-4356}
\href{https://doi.org/10.1080/01419870.2021.1977366}{doi:\nolinkurl{10.1080/01419870.2021.1977366}}


\bibitem[Memarian and Doleck(2024)]%
        {memarian_employment_2024}
\bibfield{author}{\bibinfo{person}{Bahar Memarian} {and} \bibinfo{person}{Tenzin Doleck}.} \bibinfo{year}{2024}\natexlab{}.
\newblock \showarticletitle{({Un})employment of {AI} in {Higher} {Education}}.
\newblock \bibinfo{journal}{\emph{SN Computer Science}} \bibinfo{volume}{5}, \bibinfo{number}{7} (\bibinfo{date}{Aug.} \bibinfo{year}{2024}), \bibinfo{pages}{812}.
\newblock
\showISSN{2661-8907}
\href{https://doi.org/10.1007/s42979-024-03138-z}{doi:\nolinkurl{10.1007/s42979-024-03138-z}}


\bibitem[Monteiro and Haan(2022)]%
        {monteiro_life_2022}
\bibfield{author}{\bibinfo{person}{Laura Monteiro} {and} \bibinfo{person}{Michael Haan}.} \bibinfo{year}{2022}\natexlab{}.
\newblock \showarticletitle{The {Life} {Satisfaction} of {Immigrants} in {Canada}: {Does} {Time} {Since} {Arrival} {Matter} more than {Income}?}
\newblock \bibinfo{journal}{\emph{Journal of International Migration and Integration}} \bibinfo{volume}{23}, \bibinfo{number}{3} (\bibinfo{date}{Sept.} \bibinfo{year}{2022}), \bibinfo{pages}{1397--1420}.
\newblock
\showISSN{1488-3473, 1874-6365}
\href{https://doi.org/10.1007/s12134-021-00899-x}{doi:\nolinkurl{10.1007/s12134-021-00899-x}}


\bibitem[Moran and Ackerman(2024)]%
        {moran_can_2024}
\bibfield{author}{\bibinfo{person}{Nora Moran} {and} \bibinfo{person}{David Ackerman}.} \bibinfo{year}{2024}\natexlab{}.
\newblock \showarticletitle{“{Can} {AI} really help me land a job?” {Student} reactions to the use of artificial intelligence in career preparation}.
\newblock \bibinfo{journal}{\emph{Journal of Education for Business}} \bibinfo{volume}{99}, \bibinfo{number}{2} (\bibinfo{date}{Feb.} \bibinfo{year}{2024}), \bibinfo{pages}{103--112}.
\newblock
\showISSN{0883-2323, 1940-3356}
\href{https://doi.org/10.1080/08832323.2023.2275205}{doi:\nolinkurl{10.1080/08832323.2023.2275205}}


\bibitem[Morandini et~al\mbox{.}(2023)]%
        {morandini_impact_2023}
\bibfield{author}{\bibinfo{person}{Sofia Morandini}, \bibinfo{person}{Federico Fraboni}, \bibinfo{person}{Marco De~Angelis}, \bibinfo{person}{Gabriele Puzzo}, \bibinfo{person}{Davide Giusino}, {and} \bibinfo{person}{Luca Pietrantoni}.} \bibinfo{year}{2023}\natexlab{}.
\newblock \showarticletitle{The {Impact} of {Artificial} {Intelligence} on {Workers}’ {Skills}: {Upskilling} and {Reskilling} in {Organisations}}.
\newblock \bibinfo{journal}{\emph{Informing Science: The International Journal of an Emerging Transdiscipline}}  \bibinfo{volume}{26} (\bibinfo{year}{2023}), \bibinfo{pages}{039--068}.
\newblock
\showISSN{1547-9684, 1521-4672}
\href{https://doi.org/10.28945/5078}{doi:\nolinkurl{10.28945/5078}}


\bibitem[{Natural Sciences and Engineering Research Council of Canada (NSERC)}(2025)]%
        {NSERC_USRA_2025}
\bibfield{author}{\bibinfo{person}{{Natural Sciences and Engineering Research Council of Canada (NSERC)}}.} \bibinfo{year}{2025}\natexlab{}.
\newblock \bibinfo{title}{Undergraduate Student Research Awards}.
\newblock \bibinfo{howpublished}{\url{https://nserc-crsng.canada.ca/en/funding-opportunity/undergraduate-student-research-awards}}.
\newblock
\newblock
\shownote{Accessed: 2026-01-11}.


\bibitem[Nigar et~al\mbox{.}(2025)]%
        {nigar_artificial_2025}
\bibfield{author}{\bibinfo{person}{Meher Nigar}, \bibinfo{person}{Jannatul~Ferdous Juli}, \bibinfo{person}{Uttam Golder}, \bibinfo{person}{Mohammad~Jahangir Alam}, {and} \bibinfo{person}{Mohammad~Kamal Hossain}.} \bibinfo{year}{2025}\natexlab{}.
\newblock \showarticletitle{Artificial intelligence and technological unemployment: {Understanding} trends, technology's adverse roles, and current mitigation guidelines}.
\newblock \bibinfo{journal}{\emph{Journal of Open Innovation: Technology, Market, and Complexity}} \bibinfo{volume}{11}, \bibinfo{number}{3} (\bibinfo{date}{Sept.} \bibinfo{year}{2025}), \bibinfo{pages}{100607}.
\newblock
\showISSN{21998531}
\href{https://doi.org/10.1016/j.joitmc.2025.100607}{doi:\nolinkurl{10.1016/j.joitmc.2025.100607}}


\bibitem[Norris(2024)]%
        {norris_ai_2024}
\bibfield{author}{\bibinfo{person}{Trevor Norris}.} \bibinfo{year}{2024}\natexlab{}.
\newblock \showarticletitle{{AI} {Is} the {Perfect} {Student}: {Anthropocene}, {Technocene}, and the {Ends} of {Intelligence}}.
\newblock \bibinfo{journal}{\emph{Brock Education Journal}} \bibinfo{volume}{33}, \bibinfo{number}{3} (\bibinfo{date}{Aug.} \bibinfo{year}{2024}).
\newblock
\showISSN{2371-7750, 1183-1189}
\href{https://doi.org/10.26522/brocked.v33i3.1172}{doi:\nolinkurl{10.26522/brocked.v33i3.1172}}


\bibitem[Nowrozy(2024)]%
        {nowrozy_gpts_2024}
\bibfield{author}{\bibinfo{person}{Raza Nowrozy}.} \bibinfo{year}{2024}\natexlab{}.
\newblock \showarticletitle{{GPTs} or {Grim} {Position} {Threats}? {The} {Potential} {Impacts} of {Large} {Language} {Models} on {Non}-{Managerial} {Jobs} and {Certifications} in {Cybersecurity}}.
\newblock \bibinfo{journal}{\emph{Informatics}} \bibinfo{volume}{11}, \bibinfo{number}{3} (\bibinfo{date}{July} \bibinfo{year}{2024}), \bibinfo{pages}{45}.
\newblock
\showISSN{2227-9709}
\href{https://doi.org/10.3390/informatics11030045}{doi:\nolinkurl{10.3390/informatics11030045}}


\bibitem[O'Sullivan(2025)]%
        {osullivan_these_2025}
\bibfield{author}{\bibinfo{person}{Isobel O'Sullivan}.} \bibinfo{year}{2025}\natexlab{}.
\newblock \showarticletitle{These {Companies} {Have} {Already} {Replaced} {Workers} with {AI} in 2025 and 2026}.
\newblock \bibinfo{journal}{\emph{Tech.co}} (\bibinfo{date}{Dec.} \bibinfo{year}{2025}).
\newblock
\urldef\tempurl%
\url{https://tech.co/news/companies-replace-workers-with-ai}
\showURL{%
\tempurl}


\bibitem[Pandya and Wang(2024)]%
        {pandya_artificial_2024}
\bibfield{author}{\bibinfo{person}{Shyamal~S. Pandya} {and} \bibinfo{person}{Jia Wang}.} \bibinfo{year}{2024}\natexlab{}.
\newblock \showarticletitle{Artificial intelligence in career development: a scoping review}.
\newblock \bibinfo{journal}{\emph{Human Resource Development International}} \bibinfo{volume}{27}, \bibinfo{number}{3} (\bibinfo{date}{May} \bibinfo{year}{2024}), \bibinfo{pages}{324--344}.
\newblock
\showISSN{1367-8868, 1469-8374}
\href{https://doi.org/10.1080/13678868.2024.2336881}{doi:\nolinkurl{10.1080/13678868.2024.2336881}}


\bibitem[Press(2024)]%
        {press_how_2024}
\bibfield{author}{\bibinfo{person}{Alex Press}.} \bibinfo{year}{2024}\natexlab{}.
\newblock \showarticletitle{How the {U}.{S}. {Labor} {Movement} {Is} {Confronting} {AI}}.
\newblock \bibinfo{journal}{\emph{New Labor Forum}} \bibinfo{volume}{33}, \bibinfo{number}{3} (\bibinfo{date}{Sept.} \bibinfo{year}{2024}), \bibinfo{pages}{15--22}.
\newblock
\showISSN{1095-7960, 1557-2978}
\href{https://doi.org/10.1177/10957960241276516}{doi:\nolinkurl{10.1177/10957960241276516}}


\bibitem[Qin et~al\mbox{.}(2024)]%
        {qin_artificial_2024}
\bibfield{author}{\bibinfo{person}{Meng Qin}, \bibinfo{person}{Yue Wan}, \bibinfo{person}{Junyi Dou}, {and} \bibinfo{person}{Chi~Wei Su}.} \bibinfo{year}{2024}\natexlab{}.
\newblock \showarticletitle{Artificial {Intelligence}: {Intensifying} or mitigating unemployment?}
\newblock \bibinfo{journal}{\emph{Technology in Society}}  \bibinfo{volume}{79} (\bibinfo{date}{Dec.} \bibinfo{year}{2024}), \bibinfo{pages}{102755}.
\newblock
\showISSN{0160791X}
\href{https://doi.org/10.1016/j.techsoc.2024.102755}{doi:\nolinkurl{10.1016/j.techsoc.2024.102755}}


\bibitem[Ravšelj et~al\mbox{.}(2025)]%
        {ravselj_higher_2025}
\bibfield{author}{\bibinfo{person}{Dejan Ravšelj}, \bibinfo{person}{Damijana Keržič}, \bibinfo{person}{Nina Tomaževič}, \bibinfo{person}{Lan Umek}, \bibinfo{person}{Nejc Brezovar}, \bibinfo{person}{Noorminshah A.~Iahad}, \bibinfo{person}{Ali~Abdulla Abdulla}, \bibinfo{person}{Anait Akopyan}, \bibinfo{person}{Magdalena~Waleska Aldana~Segura}, \bibinfo{person}{Jehan AlHumaid}, \bibinfo{person}{Mohamed~Farouk Allam}, \bibinfo{person}{Maria Alló}, \bibinfo{person}{Raphael Papa~Kweku Andoh}, \bibinfo{person}{Octavian Andronic}, \bibinfo{person}{Yarhands~Dissou Arthur}, \bibinfo{person}{Fatih Aydın}, \bibinfo{person}{Amira Badran}, \bibinfo{person}{Roxana Balbontín-Alvarado}, \bibinfo{person}{Helmi Ben~Saad}, \bibinfo{person}{Andrea Bencsik}, \bibinfo{person}{Isaac Benning}, \bibinfo{person}{Adrian Besimi}, \bibinfo{person}{Denilson Da~Silva Bezerra}, \bibinfo{person}{Chiara Buizza}, \bibinfo{person}{Roberto Burro}, \bibinfo{person}{Anthony Bwalya}, \bibinfo{person}{Cristina Cachero},
  \bibinfo{person}{Patricia Castillo-Briceno}, \bibinfo{person}{Harold Castro}, \bibinfo{person}{Ching~Sing Chai}, \bibinfo{person}{Constadina Charalambous}, \bibinfo{person}{Thomas K.~F. Chiu}, \bibinfo{person}{Otilia Clipa}, \bibinfo{person}{Ruggero Colombari}, \bibinfo{person}{Luis José~H. Corral~Escobedo}, \bibinfo{person}{Elísio Costa}, \bibinfo{person}{Radu~George Crețulescu}, \bibinfo{person}{Marta Crispino}, \bibinfo{person}{Nicola Cucari}, \bibinfo{person}{Fergus Dalton}, \bibinfo{person}{Meva Demir~Kaya}, \bibinfo{person}{Ivo Dumić-Čule}, \bibinfo{person}{Diena Dwidienawati}, \bibinfo{person}{Ryan Ebardo}, \bibinfo{person}{Daniel~Lawer Egbenya}, \bibinfo{person}{MoezAlIslam~Ezzat Faris}, \bibinfo{person}{Miroslav Fečko}, \bibinfo{person}{Paulo Ferrinho}, \bibinfo{person}{Adrian Florea}, \bibinfo{person}{Chun~Yuen Fong}, \bibinfo{person}{Zoë Francis}, \bibinfo{person}{Alberto Ghilardi}, \bibinfo{person}{Belinka González-Fernández}, \bibinfo{person}{Daniela Hau}, \bibinfo{person}{Md.~Shamim
  Hossain}, \bibinfo{person}{Theo Hug}, \bibinfo{person}{Fany Inasius}, \bibinfo{person}{Maryam~Jaffar Ismail}, \bibinfo{person}{Hatidža Jahić}, \bibinfo{person}{Morrison~Omokiniovo Jessa}, \bibinfo{person}{Marika Kapanadze}, \bibinfo{person}{Sujita~Kumar Kar}, \bibinfo{person}{Elham~Talib Kateeb}, \bibinfo{person}{Feridun Kaya}, \bibinfo{person}{Hanaa~Ouda Khadri}, \bibinfo{person}{Masao Kikuchi}, \bibinfo{person}{Vitaliy~Mykolayovych Kobets}, \bibinfo{person}{Katerina~Metodieva Kostova}, \bibinfo{person}{Evita Krasmane}, \bibinfo{person}{Jesus Lau}, \bibinfo{person}{Wai Him~Crystal Law}, \bibinfo{person}{Florin Lazăr}, \bibinfo{person}{Lejla Lazović-Pita}, \bibinfo{person}{Vivian Wing~Yan Lee}, \bibinfo{person}{Jingtai Li}, \bibinfo{person}{Diego~Vinicio López-Aguilar}, \bibinfo{person}{Adrian Luca}, \bibinfo{person}{Ruth~Garcia Luciano}, \bibinfo{person}{Juan~D. Machin-Mastromatteo}, \bibinfo{person}{Marwa Madi}, \bibinfo{person}{Alexandre~Lourenço Manguele}, \bibinfo{person}{Rubén~Francisco
  Manrique}, \bibinfo{person}{Thumah Mapulanga}, \bibinfo{person}{Frederic Marimon}, \bibinfo{person}{Galia~Ilieva Marinova}, \bibinfo{person}{Marta Mas-Machuca}, \bibinfo{person}{Oliva Mejía-Rodríguez}, \bibinfo{person}{Maria Meletiou-Mavrotheris}, \bibinfo{person}{Silvia~Mariela Méndez-Prado}, \bibinfo{person}{José~Manuel Meza-Cano}, \bibinfo{person}{Evija Mirķe}, \bibinfo{person}{Alpana Mishra}, \bibinfo{person}{Ondrej Mital}, \bibinfo{person}{Cristina Mollica}, \bibinfo{person}{Daniel~Ionel Morariu}, \bibinfo{person}{Natalia Mospan}, \bibinfo{person}{Angel Mukuka}, \bibinfo{person}{Silvana~Guadalupe Navarro~Jiménez}, \bibinfo{person}{Irena Nikaj}, \bibinfo{person}{Maria~Mihaylova Nisheva}, \bibinfo{person}{Efi Nisiforou}, \bibinfo{person}{Joseph Njiku}, \bibinfo{person}{Singhanat Nomnian}, \bibinfo{person}{Lulzime Nuredini-Mehmedi}, \bibinfo{person}{Ernest Nyamekye}, \bibinfo{person}{Alka Obadić}, \bibinfo{person}{Abdelmohsen~Hamed Okela}, \bibinfo{person}{Dorit Olenik-Shemesh},
  \bibinfo{person}{Izabela Ostoj}, \bibinfo{person}{Kevin~Javier Peralta-Rizzo}, \bibinfo{person}{Almir Peštek}, \bibinfo{person}{Amila Pilav-Velić}, \bibinfo{person}{Dilma Rosanda~Miranda Pires}, \bibinfo{person}{Eyal Rabin}, \bibinfo{person}{Daniela Raccanello}, \bibinfo{person}{Agustine Ramie}, \bibinfo{person}{Md. Mamun~Ur Rashid}, \bibinfo{person}{Robert A.~P. Reuter}, \bibinfo{person}{Valentina Reyes}, \bibinfo{person}{Ana~Sofia Rodrigues}, \bibinfo{person}{Paul Rodway}, \bibinfo{person}{Silvia Ručinská}, \bibinfo{person}{Shorena Sadzaglishvili}, \bibinfo{person}{Ashraf Atta M.~S. Salem}, \bibinfo{person}{Gordana Savić}, \bibinfo{person}{Astrid Schepman}, \bibinfo{person}{Samia~Mokhtar Shahpo}, \bibinfo{person}{Abdelmajid Snouber}, \bibinfo{person}{Emma Soler}, \bibinfo{person}{Bengi Sonyel}, \bibinfo{person}{Eliza Stefanova}, \bibinfo{person}{Anna Stone}, \bibinfo{person}{Artur Strzelecki}, \bibinfo{person}{Tetsuji Tanaka}, \bibinfo{person}{Carolina Tapia~Cortes}, \bibinfo{person}{Andrea
  Teira-Fachado}, \bibinfo{person}{Henri Tilga}, \bibinfo{person}{Jelena Titko}, \bibinfo{person}{Maryna Tolmach}, \bibinfo{person}{Dedi Turmudi}, \bibinfo{person}{Laura Varela-Candamio}, \bibinfo{person}{Ioanna Vekiri}, \bibinfo{person}{Giada Vicentini}, \bibinfo{person}{Erisher Woyo}, \bibinfo{person}{Özlem Yorulmaz}, \bibinfo{person}{Said A.~S. Yunus}, \bibinfo{person}{Ana-Maria Zamfir}, \bibinfo{person}{Munyaradzi Zhou}, {and} \bibinfo{person}{Aleksander Aristovnik}.} \bibinfo{year}{2025}\natexlab{}.
\newblock \showarticletitle{Higher education students’ perceptions of {ChatGPT}: {A} global study of early reactions}.
\newblock \bibinfo{journal}{\emph{PLOS ONE}} \bibinfo{volume}{20}, \bibinfo{number}{2} (\bibinfo{date}{Feb.} \bibinfo{year}{2025}), \bibinfo{pages}{e0315011}.
\newblock
\showISSN{1932-6203}
\href{https://doi.org/10.1371/journal.pone.0315011}{doi:\nolinkurl{10.1371/journal.pone.0315011}}


\bibitem[Rehak(2025)]%
        {rehak_ai_2025}
\bibfield{author}{\bibinfo{person}{Rainer Rehak}.} \bibinfo{year}{2025}\natexlab{}.
\newblock \showarticletitle{{AI} {Narrative} {Breakdown}. {A} {Critical} {Assessment} of {Power} and {Promise}}. In \bibinfo{booktitle}{\emph{Proceedings of the 2025 {ACM} {Conference} on {Fairness}, {Accountability}, and {Transparency}}}. \bibinfo{publisher}{ACM}, \bibinfo{address}{Athens Greece}, \bibinfo{pages}{1250--1260}.
\newblock
\showISBNx{979-8-4007-1482-5}
\href{https://doi.org/10.1145/3715275.3732083}{doi:\nolinkurl{10.1145/3715275.3732083}}


\bibitem[Rizun et~al\mbox{.}(2024)]%
        {rizun_employment_2024}
\bibfield{author}{\bibinfo{person}{Nina Rizun}, \bibinfo{person}{Halina Ryzhkova}, \bibinfo{person}{Irena Pawlyszyn}, {and} \bibinfo{person}{Charalampos Alexopoulos}.} \bibinfo{year}{2024}\natexlab{}.
\newblock \showarticletitle{Employment of {University} {Graduates} in the {Era} of {Digitalization} and {Artificial} {Intelligence}: {Challenges} and {Prospects}}. In \bibinfo{booktitle}{\emph{Proceedings of the 25th {Annual} {International} {Conference} on {Digital} {Government} {Research}}}. \bibinfo{publisher}{ACM}, \bibinfo{address}{Taipei Taiwan}, \bibinfo{pages}{1037--1039}.
\newblock
\showISBNx{979-8-4007-0988-3}
\href{https://doi.org/10.1145/3657054.3659123}{doi:\nolinkurl{10.1145/3657054.3659123}}


\bibitem[Sakib et~al\mbox{.}(2023)]%
        {sakib_chatgpt_2023}
\bibfield{author}{\bibinfo{person}{Nazmus Sakib}, \bibinfo{person}{Fahim~Islam Anik}, {and} \bibinfo{person}{Lei Li}.} \bibinfo{year}{2023}\natexlab{}.
\newblock \showarticletitle{{ChatGPT} in {IT} {Education} {Ecosystem}: {Unraveling} {Long}-{Term} {Impacts} on {Job} {Market}, {Student} {Learning}, and {Ethical} {Practices}}. In \bibinfo{booktitle}{\emph{The 24th {Annual} {Conference} on {Information} {Technology} {Education}}}. \bibinfo{publisher}{ACM}, \bibinfo{address}{Marietta GA USA}, \bibinfo{pages}{73--78}.
\newblock
\showISBNx{979-8-4007-0130-6}
\href{https://doi.org/10.1145/3585059.3611447}{doi:\nolinkurl{10.1145/3585059.3611447}}


\bibitem[Sallam et~al\mbox{.}(2024)]%
        {sallam_anxiety_2024}
\bibfield{author}{\bibinfo{person}{Malik Sallam}, \bibinfo{person}{Kholoud Al-Mahzoum}, \bibinfo{person}{Yousef~Meteb Almutairi}, \bibinfo{person}{Omar Alaqeel}, \bibinfo{person}{Anan Abu~Salami}, \bibinfo{person}{Zaid~Elhab Almutairi}, \bibinfo{person}{Alhur~Najem Alsarraf}, {and} \bibinfo{person}{Muna Barakat}.} \bibinfo{year}{2024}\natexlab{}.
\newblock \showarticletitle{Anxiety among {Medical} {Students} {Regarding} {Generative} {Artificial} {Intelligence} {Models}: {A} {Pilot} {Descriptive} {Study}}.
\newblock \bibinfo{journal}{\emph{International Medical Education}} \bibinfo{volume}{3}, \bibinfo{number}{4} (\bibinfo{date}{Oct.} \bibinfo{year}{2024}), \bibinfo{pages}{406--425}.
\newblock
\showISSN{2813-141X}
\href{https://doi.org/10.3390/ime3040031}{doi:\nolinkurl{10.3390/ime3040031}}


\bibitem[Sestino and De~Mauro(2022)]%
        {sestino_leveraging_2022}
\bibfield{author}{\bibinfo{person}{Andrea Sestino} {and} \bibinfo{person}{Andrea De~Mauro}.} \bibinfo{year}{2022}\natexlab{}.
\newblock \showarticletitle{Leveraging {Artificial} {Intelligence} in {Business}: {Implications}, {Applications} and {Methods}}.
\newblock \bibinfo{journal}{\emph{Technology Analysis \& Strategic Management}} \bibinfo{volume}{34}, \bibinfo{number}{1} (\bibinfo{date}{Jan.} \bibinfo{year}{2022}), \bibinfo{pages}{16--29}.
\newblock
\showISSN{0953-7325, 1465-3990}
\href{https://doi.org/10.1080/09537325.2021.1883583}{doi:\nolinkurl{10.1080/09537325.2021.1883583}}


\bibitem[Sharma et~al\mbox{.}(2025)]%
        {sharma_psychological_2025}
\bibfield{author}{\bibinfo{person}{Vinod Sharma}, \bibinfo{person}{Saikat Deb}, \bibinfo{person}{Yogesh Mahajan}, \bibinfo{person}{Avishek Ghosal}, {and} \bibinfo{person}{Manohar Kapse}.} \bibinfo{year}{2025}\natexlab{}.
\newblock \showarticletitle{Psychological impacts of {AI}-induced job displacement among {Indian} {IT} professionals: a {Delphi}-validated thematic analysis}.
\newblock \bibinfo{journal}{\emph{International Journal of Qualitative Studies on Health and Well-being}} \bibinfo{volume}{20}, \bibinfo{number}{1} (\bibinfo{date}{Dec.} \bibinfo{year}{2025}), \bibinfo{pages}{2556445}.
\newblock
\showISSN{1748-2631}
\href{https://doi.org/10.1080/17482631.2025.2556445}{doi:\nolinkurl{10.1080/17482631.2025.2556445}}


\bibitem[Shimray and Subaveerapandiyan(2025)]%
        {shimray_ai_2025}
\bibfield{author}{\bibinfo{person}{Somipam~R. Shimray} {and} \bibinfo{person}{A. Subaveerapandiyan}.} \bibinfo{year}{2025}\natexlab{}.
\newblock \showarticletitle{{AI} and workforce dynamics: a bibliometric analysis of job creation, displacement and reskilling}.
\newblock \bibinfo{journal}{\emph{Global Knowledge, Memory and Communication}} (\bibinfo{date}{Sept.} \bibinfo{year}{2025}).
\newblock
\showISSN{2514-9342, 2514-9350}
\href{https://doi.org/10.1108/GKMC-12-2024-0824}{doi:\nolinkurl{10.1108/GKMC-12-2024-0824}}


\bibitem[Si et~al\mbox{.}(2024)]%
        {si_can_2024}
\bibfield{author}{\bibinfo{person}{Chenglei Si}, \bibinfo{person}{Diyi Yang}, {and} \bibinfo{person}{Tatsunori Hashimoto}.} \bibinfo{year}{2024}\natexlab{}.
\newblock \bibinfo{title}{Can {LLMs} {Generate} {Novel} {Research} {Ideas}? {A} {Large}-{Scale} {Human} {Study} with 100+ {NLP} {Researchers}}.
\newblock
\href{https://doi.org/10.48550/arXiv.2409.04109}{doi:\nolinkurl{10.48550/arXiv.2409.04109}}
\newblock
\shownote{arXiv:2409.04109 [cs]}.


\bibitem[{Statistics Canada}({[n.\,d.]})]%
        {statistics_canada_postsecondary_nodate}
\bibfield{author}{\bibinfo{person}{{Statistics Canada}}.} \bibinfo{year}{[n.\,d.]}\natexlab{}.
\newblock \bibinfo{title}{Postsecondary enrolments, by {International} {Standard} {Classification} of {Education}, institution type, {Classification} of {Instructional} {Programs}, {STEM} and {BHASE} groupings, status of student in {Canada}, age group and gender}.
\newblock
\href{https://doi.org/10.25318/3710016301-ENG}{doi:\nolinkurl{10.25318/3710016301-ENG}}


\bibitem[{Statistics Canada}(2023)]%
        {statistics_canada_immigrants_2023}
\bibfield{author}{\bibinfo{person}{{Statistics Canada}}.} \bibinfo{year}{2023}\natexlab{}.
\newblock \showarticletitle{Immigrants’ sense of belonging to {Canada} by province of residence}.
\newblock  (\bibinfo{year}{2023}).
\newblock
\href{https://doi.org/10.25318/36280001202300600003-ENG}{doi:\nolinkurl{10.25318/36280001202300600003-ENG}}
\newblock
\shownote{Publisher: Government of Canada}.


\bibitem[{Statistics Canada}(2025)]%
        {statistics_canada_canadian_2025}
\bibfield{author}{\bibinfo{person}{{Statistics Canada}}.} \bibinfo{year}{2025}\natexlab{}.
\newblock \bibinfo{title}{Canadian postsecondary enrolments and graduates, 2023/2024}.
\newblock \bibinfo{howpublished}{The Daily}.
\newblock
\urldef\tempurl%
\url{https://www150.statcan.gc.ca/n1/daily-quotidien/251120/dq251120e-eng.htm}
\showURL{%
\tempurl}


\bibitem[{Statistics Canada}(2026)]%
        {statistics_canada_labour_2026}
\bibfield{author}{\bibinfo{person}{{Statistics Canada}}.} \bibinfo{year}{2026}\natexlab{}.
\newblock \bibinfo{title}{Labour {Force} {Survey}, {December} 2025}.
\newblock \bibinfo{howpublished}{The Daily}.
\newblock
\urldef\tempurl%
\url{https://www150.statcan.gc.ca/n1/daily-quotidien/260109/dq260109a-eng.htm}
\showURL{%
\tempurl}


\bibitem[Sun(2021)]%
        {sun_research_2021}
\bibfield{author}{\bibinfo{person}{Tingyu Sun}.} \bibinfo{year}{2021}\natexlab{}.
\newblock \showarticletitle{Research on the {Influence} of {Artificial} {Intelligence} on {Female} {Labor} {Employment}}. In \bibinfo{booktitle}{\emph{2021 3rd {International} {Conference} on {Artificial} {Intelligence} and {Advanced} {Manufacture}}}. \bibinfo{publisher}{ACM}, \bibinfo{address}{Manchester United Kingdom}, \bibinfo{pages}{2805--2809}.
\newblock
\showISBNx{978-1-4503-8504-6}
\href{https://doi.org/10.1145/3495018.3501186}{doi:\nolinkurl{10.1145/3495018.3501186}}


\bibitem[Tereshchenko et~al\mbox{.}(2024)]%
        {tereshchenko_why_2024}
\bibfield{author}{\bibinfo{person}{Elizaveta Tereshchenko}, \bibinfo{person}{Sherah Kurnia}, \bibinfo{person}{Audrey Mbogho}, {and} \bibinfo{person}{Ari Happonen}.} \bibinfo{year}{2024}\natexlab{}.
\newblock \showarticletitle{Why do {Women} {Refrain} from {IT}/{ICT} studies at {Higher} {Education} {Institutions}? {A} {Literature} {Review}}.
\newblock \bibinfo{journal}{\emph{International Journal of Gender, Science and Technology}} \bibinfo{volume}{16}, \bibinfo{number}{2} (\bibinfo{date}{Aug.} \bibinfo{year}{2024}), \bibinfo{pages}{101--123}.
\newblock
\urldef\tempurl%
\url{https://genderandset.open.ac.uk/index.php/genderandset/article/view/1297}
\showURL{%
\tempurl}
\newblock
\shownote{Section: Research and theoretical papers}.


\bibitem[Tomar et~al\mbox{.}(2024)]%
        {tomar_future_2024}
\bibfield{author}{\bibinfo{person}{Anjali Tomar}, \bibinfo{person}{{Arti}}, \bibinfo{person}{Simran Suman}, {and} \bibinfo{person}{Seema Sharma}.} \bibinfo{year}{2024}\natexlab{}.
\newblock \showarticletitle{The {Future} of {Work}: {Impacts} of {AI} on {Employment} and {Job} {Market} {Dynamics}}. In \bibinfo{booktitle}{\emph{2024 {International} {Conference} on {Progressive} {Innovations} in {Intelligent} {Systems} and {Data} {Science} ({ICPIDS})}}. \bibinfo{publisher}{IEEE}, \bibinfo{address}{Pattaya, Thailand}, \bibinfo{pages}{181--184}.
\newblock
\showISBNx{979-8-3315-3469-1}
\href{https://doi.org/10.1109/ICPIDS65698.2024.00037}{doi:\nolinkurl{10.1109/ICPIDS65698.2024.00037}}


\bibitem[Wang et~al\mbox{.}(2024)]%
        {wang_investigating_2024}
\bibfield{author}{\bibinfo{person}{Ruotong Wang}, \bibinfo{person}{Ruijia Cheng}, \bibinfo{person}{Denae Ford}, {and} \bibinfo{person}{Thomas Zimmermann}.} \bibinfo{year}{2024}\natexlab{}.
\newblock \showarticletitle{Investigating and {Designing} for {Trust} in {AI}-powered {Code} {Generation} {Tools}}. In \bibinfo{booktitle}{\emph{The 2024 {ACM} {Conference} on {Fairness}, {Accountability}, and {Transparency}}}. \bibinfo{publisher}{ACM}, \bibinfo{address}{Rio de Janeiro Brazil}, \bibinfo{pages}{1475--1493}.
\newblock
\showISBNx{979-8-4007-0450-5}
\href{https://doi.org/10.1145/3630106.3658984}{doi:\nolinkurl{10.1145/3630106.3658984}}


\bibitem[Wang et~al\mbox{.}(2023)]%
        {wang_exploring_2023}
\bibfield{author}{\bibinfo{person}{Ting Wang}, \bibinfo{person}{Brady~D. Lund}, \bibinfo{person}{Agostino Marengo}, \bibinfo{person}{Alessandro Pagano}, \bibinfo{person}{Nishith~Reddy Mannuru}, \bibinfo{person}{Zoë~A. Teel}, {and} \bibinfo{person}{Jenny Pange}.} \bibinfo{year}{2023}\natexlab{}.
\newblock \showarticletitle{Exploring the {Potential} {Impact} of {Artificial} {Intelligence} ({AI}) on {International} {Students} in {Higher} {Education}: {Generative} {AI}, {Chatbots}, {Analytics}, and {International} {Student} {Success}}.
\newblock \bibinfo{journal}{\emph{Applied Sciences}} \bibinfo{volume}{13}, \bibinfo{number}{11} (\bibinfo{date}{May} \bibinfo{year}{2023}), \bibinfo{pages}{6716}.
\newblock
\showISSN{2076-3417}
\href{https://doi.org/10.3390/app13116716}{doi:\nolinkurl{10.3390/app13116716}}


\bibitem[Wang et~al\mbox{.}(2022)]%
        {wang_what_2022}
\bibfield{author}{\bibinfo{person}{Yu-Min Wang}, \bibinfo{person}{Chung-Lun Wei}, \bibinfo{person}{Hsin-Hui Lin}, \bibinfo{person}{Sheng-Ching Wang}, {and} \bibinfo{person}{Yi-Shun Wang}.} \bibinfo{year}{2022}\natexlab{}.
\newblock \showarticletitle{What drives students’ {AI} learning behavior: a perspective of {AI} anxiety}.
\newblock \bibinfo{journal}{\emph{Interactive Learning Environments}} (\bibinfo{date}{Dec.} \bibinfo{year}{2022}), \bibinfo{pages}{1--17}.
\newblock
\showISSN{1049-4820, 1744-5191}
\href{https://doi.org/10.1080/10494820.2022.2153147}{doi:\nolinkurl{10.1080/10494820.2022.2153147}}


\bibitem[Wei(2025)]%
        {wei_i_2025}
\bibfield{author}{\bibinfo{person}{Li-Wei Wei}.} \bibinfo{year}{2025}\natexlab{}.
\newblock \showarticletitle{“{I} {Am} {Creative}, {Therefore} {I} {Am} {Employable}”: {Self}-{Creativity} {Identity} as a {Mediator} of the {AI}-{Unemployment} {Anxiety} {Link}}.
\newblock \bibinfo{journal}{\emph{Beijing International Review of Education}} \bibinfo{volume}{7}, \bibinfo{number}{3} (\bibinfo{date}{Sept.} \bibinfo{year}{2025}), \bibinfo{pages}{224--238}.
\newblock
\showISSN{2590-2547, 2590-2539}
\href{https://doi.org/10.1177/25902547251372084}{doi:\nolinkurl{10.1177/25902547251372084}}


\bibitem[Yang et~al\mbox{.}(2026)]%
        {yang_deskilling_2026}
\bibfield{author}{\bibinfo{person}{Bo Yang}, \bibinfo{person}{Yongqiang Sun}, \bibinfo{person}{Zihan Zeng}, {and} \bibinfo{person}{Qinwei Li}.} \bibinfo{year}{2026}\natexlab{}.
\newblock \showarticletitle{Deskilling, reskilling, or upskilling? {Unpacking} the pathways of student adaptation to generative artificial intelligence}.
\newblock \bibinfo{journal}{\emph{International Journal of Information Management}}  \bibinfo{volume}{87} (\bibinfo{date}{April} \bibinfo{year}{2026}), \bibinfo{pages}{103002}.
\newblock
\showISSN{02684012}
\href{https://doi.org/10.1016/j.ijinfomgt.2025.103002}{doi:\nolinkurl{10.1016/j.ijinfomgt.2025.103002}}


\bibitem[Zhang(2025)]%
        {zhang_ai-driven_2025}
\bibfield{author}{\bibinfo{person}{Qianyi Zhang}.} \bibinfo{year}{2025}\natexlab{}.
\newblock \showarticletitle{{AI}-driven unemployment risk and household financial decision: {Evidence} from {China}}.
\newblock \bibinfo{journal}{\emph{Journal of Asian Economics}}  \bibinfo{volume}{99} (\bibinfo{date}{Aug.} \bibinfo{year}{2025}), \bibinfo{pages}{101963}.
\newblock
\showISSN{10490078}
\href{https://doi.org/10.1016/j.asieco.2025.101963}{doi:\nolinkurl{10.1016/j.asieco.2025.101963}}


\bibitem[Zhang et~al\mbox{.}(2025)]%
        {zhang_llm_2025}
\bibfield{author}{\bibinfo{person}{Ziyao Zhang}, \bibinfo{person}{Chong Wang}, \bibinfo{person}{Yanlin Wang}, \bibinfo{person}{Ensheng Shi}, \bibinfo{person}{Yuchi Ma}, \bibinfo{person}{Wanjun Zhong}, \bibinfo{person}{Jiachi Chen}, \bibinfo{person}{Mingzhi Mao}, {and} \bibinfo{person}{Zibin Zheng}.} \bibinfo{year}{2025}\natexlab{}.
\newblock \showarticletitle{{LLM} {Hallucinations} in {Practical} {Code} {Generation}: {Phenomena}, {Mechanism}, and {Mitigation}}.
\newblock \bibinfo{journal}{\emph{Proceedings of the ACM on Software Engineering}} \bibinfo{volume}{2}, \bibinfo{number}{ISSTA} (\bibinfo{date}{June} \bibinfo{year}{2025}), \bibinfo{pages}{481--503}.
\newblock
\showISSN{2994-970X}
\href{https://doi.org/10.1145/3728894}{doi:\nolinkurl{10.1145/3728894}}


\bibitem[Ziegler et~al\mbox{.}(2024)]%
        {ziegler_measuring_2024}
\bibfield{author}{\bibinfo{person}{Albert Ziegler}, \bibinfo{person}{Eirini Kalliamvakou}, \bibinfo{person}{X.~Alice Li}, \bibinfo{person}{Andrew Rice}, \bibinfo{person}{Devon Rifkin}, \bibinfo{person}{Shawn Simister}, \bibinfo{person}{Ganesh Sittampalam}, {and} \bibinfo{person}{Edward Aftandilian}.} \bibinfo{year}{2024}\natexlab{}.
\newblock \showarticletitle{Measuring {GitHub} {Copilot}'s {Impact} on {Productivity}}.
\newblock \bibinfo{journal}{\emph{Commun. ACM}} \bibinfo{volume}{67}, \bibinfo{number}{3} (\bibinfo{date}{March} \bibinfo{year}{2024}), \bibinfo{pages}{54--63}.
\newblock
\showISSN{0001-0782, 1557-7317}
\href{https://doi.org/10.1145/3633453}{doi:\nolinkurl{10.1145/3633453}}


\end{thebibliography}

\newpage
\section*{Appendices}
\label{sec:appendix}

% \subsection{Survey Questions}
% \begin{table}[h]
% \centering
% \caption{7-pt Likert scale survey questions}
% \begin{tabular}{|p{0.9\linewidth}|}
% \hline
% \begin{itemize}[leftmargin=1em]
%     \item \textit{I am stressed or anxious about my future career prospects because of AI}
%     \item \textit{I feel that my future career roles could be replaced or displaced because of AI}
%     \item \textit{I have upskilled or been motivated to upskill because of AI}
%     \item \textit{I have reskilled or been motivated to reskill because of AI}
%     \item \textit{I have selected courses I normally would not have considered taking because of AI}
%     \item \textit{I have decided against selecting courses I previously wanted to take because of AI}
%     \item \textit{I have considered leaving my current field of study or have left a field of study in the past because of AI}
%     \item \textit{I have felt discouraged from entering certain roles or industries I was once considering because of AI}
% \end{itemize}
% \\ \hline 
% \end{tabular}
% \end{table}

\subsection*{Interview Protocol}
\label{methods:protocol}

During interviews, students were invited to discuss their initial responses to the survey in greater depth, and raise additional topics of interest to them. We prepared the following questions to shape the conversation:

\begin{table}[h]
\centering
\caption{Questions outlined in interview protocol.}
\begin{tabular}{|p{0.9\linewidth}|}
\hline
\begin{enumerate}
    \item Why are (or are you not) you stressed or anxious about your future career prospects to the degree you declared?
    \item How do you use AI for school/work/research, and how dose it relate to your anxiety about future career prospects?
    \item Why have you (or have you not) been motivated to upskill due to AI?
    \item Why have you (or have you not) been motivated to reskill due to AI?
    \item Which programs have you considered enrolling in or decided against enrolling in due to AI?
    \begin{enumerate}
        \item Which courses have you taken or decided against taking due to AI?
        \item Do you find that your coursework is easily replaceable by AI?
    \end{enumerate}
    \item Why are you interested in your chosen subfields, and how does it relate to AI anxiety?
    \item How are your plans after graduation related to AI anxiety?
    \begin{enumerate}
        \item Why have you (or have you not) felt discouraged from entering roles or industries you were considering because of AI?
    \end{enumerate}
    \item Does being an international/domestic student influence your job anxiety in any way?
    \item Would anything alleviate your AI anxiety?
\end{enumerate}
\\ \hline 
\end{tabular}
\end{table}

These questions were prepared to answer our research questions; specifically, Questions 1, 2, 8, and 9 answer RQ1, Questions 6 and 7 answer RQ2, and Questions 3, 4, and 5 answer RQ3. We chose the order of the questions to shape the most natural conversation possible, beginning with why or why not the participant experiences stress or anxiety due to AI job replacement, discussing upskilling and reskilling, plans after graduation, and ending with international student struggles and what would alleviate the participant's anxiety.

Note that we reiterated the definitions of upskilling and reskilling before Questions 3 and 4. Our discussion of these terms was derived from the literature \cite{forum_future_2025}. Some questions were further developed after initial interviews. Question 5(b) was added in the third interview, which was the first interview with a first year student. Question 8 was added in the fourth interview, which was the second interview with an international student. Moreover, Question 9 was added halfway through the interview process to gauge how participants think they can mitigate their anxieties. Lastly, follow-up questions were asked depending on how substantial the participants' responses were to a particular question.

\end{document}